\documentclass[longauth]{aa} 

\usepackage{graphicx}
\usepackage{natbib}

\usepackage{booktabs}  
\usepackage[colorlinks=true,linkcolor=blue,citecolor=blue]{hyperref}
\usepackage{txfonts}

\graphicspath{{./Pictures_used/}} 

\providecommand{\abs}[1]{\lvert#1\rvert} 

\usepackage[dvipsnames]{xcolor} 

\usepackage{longtable}
\usepackage{lscape}

\usepackage{orcidlink}

\usepackage{amstext}

\begin{document}

   \title{}

   \subtitle{Two sub-Neptunes around the M dwarf TOI-1470}

   \author{E.~Gonz\'alez-\'Alvarez\,\orcidlink{0000-0002-4820-2053} \inst{\ref{inst:CSIC-INTA1}, \ref{inst:UCM}} 
   \and M.\,R.~Zapatero Osorio \inst{\ref{inst:CSIC-INTA1}}
   \and J.\,A.~Caballero\,\orcidlink{0000-0002-7349-1387} \inst{\ref{inst:CSIC-INTA1}}
   \and V.\,J.\,S.~B\'ejar \inst{\ref{inst:IAC}, \ref{inst:ULL}}
   \and C.~Cifuentes \inst{\ref{inst:CSIC-INTA1}}
   \and A.~Fukui\,\orcidlink{0000-0002-4909-5763} \inst{\ref{inst:Uni_Tokyo},\ref{inst:IAC}}
   \and E.~Herrero \inst{\ref{inst:IEEC}}
   \and K.~Kawauchi \inst{\ref{inst:Multi-disc_Tokyo}}
   \and J.\,H. Livingston\,\orcidlink{0000-0002-4881-3620} \inst{\ref{inst:Astrob_Tokyo}, \ref{inst:National_astro_obs_Tokyo}, \ref{inst:Dep_astro_science_Tokyo}}
   \and M.\,J.~L\'opez-Gonz\'alez \inst{\ref{inst:IAA}}
   \and G.~Morello \inst{\ref{inst:IAC}, \ref{inst:ULL}}
   \and F.~Murgas \inst{\ref{inst:IAC},\ref{inst:ULL}}
   \and N.~ Narita \inst{\ref{inst:Uni_Tokyo},\ref{inst:Astrob_Tokyo}, \ref{inst:IAC}}
   \and E.~Pall\'e \inst{\ref{inst:IAC},\ref{inst:ULL}}
   \and V.\,M.~Passegger \inst{\ref{inst:Hamb}, \ref{inst:USA},\ref{inst:IAC}, \ref{inst:ULL}}
   \and E.~Rodr\'iguez \inst{\ref{inst:IAA}}
   \and C.~Rodr\'iguez-L\'opez \inst{\ref{inst:IAA}}
   \and J.~Sanz-Forcada\inst{\ref{inst:CSIC-INTA1}}
   \and A.~Schweitzer \inst{\ref{inst:Hamb}}
   \and H.\,M.~Tabernero \inst{\ref{inst:CSIC-INTA1}}
   \and A.~Quirrenbach \inst{\ref{inst:zah_lsw}}
   \and P.\,J.~Amado\inst{\ref{inst:IAA}}
   \and D.~Charbonneau\,\orcidlink{0000-0002-9003-484X} \inst{\ref{inst:Harv&Smith}}
   \and D.\,R.~Ciardi \inst{\ref{inst:NASA_inst}}
   \and S.~Cikota \inst{\ref{inst:CAHA}}
   \and K.\,A.~Collins\,\orcidlink{0000-0001-6588-9574} \inst{\ref{inst:Harv&Smith}}
   \and D.\, M.~Conti\,\orcidlink{0000-0003-2239-0567} \inst{\ref{inst:American_Asso_MA}}
   \and M.~Fausnaugh\,\orcidlink{0000-0002-9113-7162} \inst{\ref{inst:Kavli_inst}}
   \and A.\,P.~Hatzes \inst{\ref{inst:TLS}} 
   \and C.~Hedges \inst{\ref{inst:NASA_MD}}
    \and Th.~Henning \inst{\ref{inst:MPI_heil}}
    \and J.\,M.~Jenkins\,\orcidlink{0000-0002-4715-9460} \inst{\ref{inst:NASA_CA}}
    \and D.\,W.~Latham\,\orcidlink{0000-0001-9911-7388} \inst{\ref{inst:Harv&Smith}}
    \and B.~Massey\,\orcidlink{0000-0001-8879-7138} \inst{\ref{inst:Landers}}
    \and D.~Moldovan \inst{\ref{inst:Google}}
   \and D.~Montes \inst{\ref{inst:UCM}}\,\orcidlink{0000-0002-7779-238X}
   \and A.~Panahi\,\orcidlink{0000-0001-5850-4373} \inst{\ref{inst:Tel_Aviv}}
   \and A.~Reiners\inst{\ref{inst:IAG_goett}}
   \and I.~Ribas \inst{\ref{inst:IEEC-CSIC},\ref{inst:IEEC}}
   \and G.\,R.~Ricker\,\orcidlink{0000-0003-2058-6662} \inst{\ref{inst:Kavli_inst}}
   \and S.~Seager\,\orcidlink{0000-0002-6892-6948} \inst{ \ref{inst:Kavli_inst},\ref{inst:Massachusetts}, \ref{inst:MIT}}
   \and A.~Shporer\,\orcidlink{0000-0002-1836-3120} \inst{\ref{inst:Kavli_inst}}
   \and G.~Srdoc \inst{\ref{inst:Obs_Croatia}}
   \and P.~Tenenbaum\,\orcidlink{0000-0002-1949-4720} \inst{\ref{inst:SETI_USA}}
   \and R.~Vanderspek\,\orcidlink{0000-0001-6763-6562} \inst{\ref{inst:Kavli_inst}}
   \and J.\,N.~Winn\,\orcidlink{0000-0002-4265-047X} \inst{\ref{inst:Princeton}}
   \and I.~Fukuda\,\orcidlink{0000-0002-9436-2891} \inst{\ref{inst:Multi-disc_Tokyo}}
   \and M.~Ikoma\,\orcidlink{0000-0002-5658-5971} \inst{\ref{inst:Division_Science_Japan}}
   \and K.~Isogai\,\orcidlink{0000-0002-6480-3799} \inst{\ref{inst:Division_Science_Japan}, \ref{inst:Multi-disc_Tokyo}}
   \and Y.~Kawai\,\orcidlink{0000-0002-0488-6297} \inst{\ref{inst:Multi-disc_Tokyo}}
   \and M.~Mori\,\orcidlink{0000-0003-1368-6593} \inst{\ref{inst:Depart_astro_Uni_Tokyo}}
   \and M.~Tamura\,\orcidlink{0000-0002-6510-0681} \inst{\ref{inst:Depart_astro_Uni_Tokyo}, \ref{inst:Astrob_Tokyo}, \ref{inst:National_astro_obs_Tokyo} }
   \and N.~Watanabe\,\orcidlink{0000-0002-7522-8195} \inst{\ref{inst:Multi-disc_Tokyo}}
          }

                        \institute{Centro de Astrobiolog\'ia (CAB), CSIC-INTA, Carretera de Ajalvir km~4, 28850 Torrej\'on de Ardoz, Madrid, Spain \label{inst:CSIC-INTA1}
                        \and Departamento de F\'isica de la Tierra y Astrof\'isica \& IPARCOS-UCM (Instituto de F\'isica de Part\'iculas y del Cosmos de la UCM), Facultad de Ciencias F\'isicas, Universidad Complutense de Madrid, 28040 Madrid, Spain \label{inst:UCM}
                        \and Instituto de Astrof\'isica de Canarias (IAC), 38200 La Laguna, Te\-ne\-ri\-fe, Spain\label{inst:IAC}
                        \and Departamento de Astrof\'isica, Universidad de La Laguna (ULL), 38206 La Laguna, Tenerife, Spain\label{inst:ULL}  
                        \and Komaba Institute for Science, The University of Tokyo, 3-8-1 Komaba, Meguro, Tokyo 153-8902, Japan \label{inst:Uni_Tokyo}
                        \and Institut d'Estudis Espacials de Catalunya (IEEC), 08034 Barcelona, Spain  \label{inst:IEEC}
                        \and Department of Multi-Disciplinary Sciences, Graduate School of Arts and Sciences, The University of Tokyo, 3-8-1 Komaba, Meguro, Tokyo 153-8902, Japan \label{inst:Multi-disc_Tokyo}
                        \and Astrobiology Center, 2-21-1 Osawa, Mitaka, Tokyo 181-8588, Japan\label{inst:Astrob_Tokyo}
                        \and National Astronomical Observatory of Japan, 2-21-1 Osawa, Mitaka, Tokyo 181-8588, Japan \label{inst:National_astro_obs_Tokyo}
                        \and Department of Astronomical Science, The Graduated University for Advanced Studies, SOKENDAI, 2-21-1, Osawa, Mitaka, Tokyo, 181-8588, Japan \label{inst:Dep_astro_science_Tokyo}
                        \and Instituto de Astrof\'isica de Andaluc\'ia (IAA-CSIC), Glorieta de la Astronom\'ia s/n, 18008 Granada, Spain \label{inst:IAA}  
                        \and Hamburger Sternwarte, Universit\"at Hamburg, Gojenbergsweg 112,
                        21029 Hamburg, Germany \label{inst:Hamb}
                        \and Homer L. Dodge Department of Physics and Astronomy, University
                        of Oklahoma, 440 West Brooks Street, Norman, OK 73019, USA \label{inst:USA}
                        \and Landessternwarte, Zentrum f\"ur Astronomie der Universit\"at Heidelberg, K\"onigstuhl 12, 69117 Heidelberg, Germany\label{inst:zah_lsw}
                        \and Center for Astrophysics \textbar \ Harvard \& Smithsonian, 60 Garden Street, Cambridge, MA 02138, USA \label{inst:Harv&Smith}
                        \and NASA Exoplanet Science Institute, Caltech/IPAC, Mail Code 100-22, 1200 E. California Blvd., Pasadena, CA 91125, USA \label{inst:NASA_inst}
                        \and Centro Astron\'omico Hispano en Andaluc\'ia CAHA, Observatorio Astron\'omico de Calar Alto, Sierra de los Filabres, 04550, G\'ergal, Spain \label{inst:CAHA}
                        \and American Association of Variable Star Observers, 185 Alewife Brook Parkway, Suite 410, Cambridge, MA 02138, USA \label{inst:American_Asso_MA}
                        \and Department of Physics and Kavli Institute for Astrophysics and Space Research, Massachusetts Institute of Technology, Cambridge, MA 02139, USA \label{inst:Kavli_inst}
                        \and Th\"uringer Landessternwarte Tautenburg, Sternwarte 5, 07778 Tautenburg, Germany\label{inst:TLS}    
                        \and NASA Goddard Space Flight Center, 8800 Greenbelt Rd, Greenbelt, MD 20771, USA \label{inst:NASA_MD}
                        \and Max-Planck-Institut f\"ur Astronomie, K\"onigstuhl 17, 69117 Heidelberg, Germany \label{inst:MPI_heil}
                        \and Intelligent Systems Division, NASA Ames Research Center, Moffett Field, CA 94035, USA \label{inst:NASA_CA}
                        \and Villa '39 Observatory, Landers, CA 92285, USA \label{inst:Landers}
                        \and Google, Cambridge, MA 02142, USA \label{inst:Google}
                        \and School of Physics and Astronomy, Tel Aviv University, Tel Aviv, 6997801, Israel \label{inst:Tel_Aviv}
                        \and Institut f\"ur Astrophysik, Georg-August-Universit\"at G\"ottingen, Friedrich-Hund-Platz 1, 37077 G\"ottingen, Germany\label{inst:IAG_goett}
                        \and Institut de Ci\`encies de l'Espai (IEEC-CSIC), Campus UAB, Carrer de Can Magrans s/n, 08193, Bellaterra, Spain \label{inst:IEEC-CSIC} 
                        \and Department of Earth, Atmospheric, and Planetary Sciences, Massachusetts Institute of Technology, Cambridge, MA 02139, USA \label{inst:Massachusetts}
                        \and Department of Aeronautics and Astronautics, Massachusetts Institute of Technology, 77 Massachusetts Avenue, Cambridge, MA 02139, USA \label{inst:MIT}
                        \and Kotizarovci Observatory, Sarsoni 90, 51216 Viskovo, Croatia \label{inst:Obs_Croatia}
                        \and SETI Institute/NASA Ames Research Center, Moffett Field, CA 94305, USA \label{inst:SETI_USA}
                        \and Department of Astrophysical Sciences, Princeton University, Princeton, NJ 08544, USA \label{inst:Princeton}
                        \and Division of Science, National Astronomical Observatory of Japan, 2-21-1 Osawa, Mitaka, Tokyo 181-8588, Japan \label{inst:Division_Science_Japan}
                        \and Department of Astronomy, Graduate School of Science, The University of Tokyo, 7-3-1 Hongo, Bunkyo-ku, Tokyo 113-0033, Japan \label{inst:Depart_astro_Uni_Tokyo}
                        }

   \offprints{Esther~Gonz\'alez-\'Alvarez \\ \email{estgon11@ucm.es}}
   \date{Received 1 March 2023 / Accepted 7 June 2023}

 
  \abstract
   {}
   {A transiting planet candidate with a sub-Neptune radius orbiting the nearby ($d$ = 51.9$\pm$0.07\,pc) M1.5\,V star TOI-1470 with a period of $\sim$2.5\,d was announced by the NASA Transiting Exoplanet Survey Satellite (\textit{TESS}), which observed the field of TOI-1470 in four different sectors. We aim to validate its planetary nature using precise radial velocities (RVs) taken with the CARMENES spectrograph.}
   {We obtained 44 RV measurements with CARMENES spanning eight months between 3 June 2020 and 17 January 2021. For a better characterization of the parent star activity, we also collected contemporaneous optical photometric observations at the Joan Or\'o and Sierra Nevada Observatories, and we retrieved archival photometry from the literature. We used ground-based photometric observations from MuSCAT and also from MuSCAT2 and MuSCAT3 to confirm the planetary transit signals. We performed a combined photometric and spectroscopic analysis by including Gaussian processes and Keplerian orbits to simultaneously account for the stellar activity and planetary signals.}
   {We estimate that TOI-1470 has a rotation period of 29$\pm$3\,d based on photometric and spectroscopic data. The combined analysis confirms the discovery of the announced transiting planet, TOI-1470\,b, with an orbital period of 2.527093$\pm$0.000003\,d, a mass of $7.32^{+1.21}_{-1.24}$\,M$_{\oplus}$, and a radius of $2.18^{+0.04}_{-0.04}$\,R$_{\oplus}$. We also discover a second transiting planet that was not announced previously by \textit{TESS}, TOI-1470\,c, with an orbital period of 18.08816$\pm$0.00006\,d, a mass of $7.24^{+2.87}_{-2.77}$\,M$_{\oplus}$, and a radius of $2.47^{+0.02}_{-0.02}$\,R$_{\oplus}$. The two planets are placed on the same side of the radius valley of M dwarfs and lie between TOI-1470 and the inner border of its habitable zone.}
   {}


\maketitle
%

\section{Introduction}
\label{Introduction}

More than 5300 exoplanets have been discovered in the past 30 years. However, not all of them are well characterized. For exoplanets that were detected with a combination of transit photometry and the spectroscopic radial velocity (RV) method, we are able to determine the mass and the radius, and consequently, their bulk density. This approach provides valuable information in the ongoing debate about the nature and origin of super-Earths and mini-Neptunes. Various studies have revealed that the radius distribution of planets slightly larger than the Earth is bimodal \citep{2017AJ....154..109F, 2018MNRAS.479.4786V, 2020AJ....160...89P, 2020AJ....160...22C}. Super-Earth planets with radii up to $\sim$1.5\,R$_{\oplus}$ are relatively common, as are mini-Neptunes in the range of $\sim$2–4\,R$_{\oplus}$. There appears to be a deficit of planets in between these sizes, however. The location of the radius gap depends on the orbital period, the planet insolation, the spectral type of the parent star, or a combination of these parameters \citep{2018MNRAS.479.4786V, 2021MNRAS.507.2154V, 2020AJ....160...22C}. It divides the planets into two different populations, namely small rocky planets and larger planets with volatile-rich envelopes \citep{2017AJ....154..109F}. 

Here, we present two transiting planets orbiting the early-M dwarf TOI-1470. The transit signal of the shorter-period planet was first detected by the NASA Transiting Exoplanet Survey Satellite \citep[\textit{TESS};][]{2015JATIS...1a4003R}, while the longer-period planet is detected in this work for the first time. Their masses and radii are derived through a joint modeling of the \textit{TESS} photometry and follow-up RV observations with the Calar Alto high-Resolution search for M dwarfs with Exoearths with Near-infrared and optical Échelle Spectrographs \citep[CARMENES][]{2014SPIE.9147E..1FQ, 2016SPIE.9908E..12Q,2018SPIE10702E..0WQ, 2020SPIE11447E..3CQ, 2018A&A...612A..49R}, together with extensive ground-based photometric follow-up observations with the MuSCAT telescopes, Joan Or\'o, and Sierra Nevada observatories. This work is part of the \textit{TESS} follow-up program within the CARMENES guaranteed-time observations (GTO) survey, which aims to validate the planetary nature of transit events detected around M dwarfs \citep{2019A&A...628A..39L, 2022A&A...664A.199L, 
2020A&A...642A.236K, 2022A&A...659A..17K, 
2020A&A...642A.173N, 2020A&A...644A.127D,
2020A&A...639A.132B, 2021A&A...650A..78B, 
2021A&A...649A.144S, 2022AJ....163..133E, 2023A&A...670A..84K}. 

All observations of TOI-1470 are presented in Section \ref{sec:observations}. In Section \ref{sec:TOI-1470} we introduce the target star (TOI-1470) and present its stellar properties, which are newly derived and and also collected from the literature. In Section \ref{sec:analysis} we provide a detailed analysis of the photometric light curves, CARMENES RVs and spectroscopic activity indicators with the main goal of determining the presence of planet candidates. The final combined model, from which the properties of the newly discovered planets orbiting TOI-1470 are derived, is given in Section \ref{sec:planet orbiting toi-1470}. A brief discussion of the implications of this finding and the conclusions of this paper appear in Sections \ref{sec:discussion} and \ref{sec:summary}, respectively.

\section{Observations}
\label{sec:observations}

\subsection{\textit{TESS} photometry}
\label{subsec:TESS photometric time serie}

        \begin{table}[]
        \begin{small}
                \centering
                \caption{Transit observations of TOI-1470.}
                \label{tab:tess_observ_TOI-1470}
                \begin{tabular}{l c c}
                        \hline
                        \hline
                        \noalign{\smallskip}
                        Telescope  & Filter  & Date\\
                        \noalign{\smallskip}    
                        \hline  
                        \noalign{\smallskip}
                        \textit{TESS} S17 & $T$ &   7 October 2019 -- 2 November 2019\\
                        \textit{TESS} S18 & $T$ &   2 November 2019 -- 27 November 2019\\
                        \textit{TESS} S24 & $T$ &   16 April 2020 -- 13 May 2020\\
            \textit{TESS} S58 & $T$ &   29 October 2022 -- 26 November 2022 \\
                        MuSCAT                                  & $z'_s$, $r' $ , $g'$  &   31 July 2020  \\
                        MuSCAT2                                 & $z'_s$, $i'$, $g'$  &   20 September 2021 \\
                        MuSCAT3                                 & $z'_s$, $i'$, $g'$ &   26 October 2021 \\               
                                
                        \noalign{\smallskip}
                        \hline
                \end{tabular}
        \end{small}
        \end{table}

The Transiting Exoplanet Survey Satellite (\textit{TESS}) is an all-sky transit survey whose principal goal is the detection of planets smaller than Neptune orbiting bright stars that can be followed-up with observations that then lead to the determination of planetary masses and atmospheric compositions. In its primary mission, \textit{TESS} conducted high-precision photometry of more than 200,000 stars over two years until 4 July 2020. All observations were made available to the community as target pixel files (TPFs) and calibrated light curves (LC). \textit{TESS} LC files include the time stamps (TBJD = BJD \text{--} 2,457,000), simple aperture photometry (SAP) fluxes, and presearch data conditioned simple aperture photometry (PDCSAP) fluxes \citep{2012PASP..124.1000S,Stumpe2012,Stumpe2014}. The SAP flux is the flux after summing the calibrated pixels within the \textit{TESS} optimal photometric aperture, and the PDCSAP flux corresponds to the SAP flux values corrected for instrumental variations and for crowding. The optimal photometric aperture is defined such that the stellar signal has a high signal-to-noise ratio, with minimum contamination from the background. The \textit{TESS} detector bandpass spans from 600 to 1000\,nm and is centered on the traditional Cousins $I$ band (786.5\,nm). This wide red optical bandpass was chosen to reduce photon-counting noise and to increase the sensitivity to small planets transiting cool red stars.

TOI-1470 (TIC 284441182) was observed by \textit{TESS} in 2 min short-cadence integrations in sectors 17, 18, 24, and 58 during the \textit{TESS} primary mission (see Table~\ref{tab:tess_observ_TOI-1470}). All sectors were processed by the Science Processing Operations Center pipeline \citep[SPOC,][]{2016SPIE.9913E..3EJ} and searched for transiting planet signatures with an adaptive wavelet-based transit detection algorithm \citep{2002ApJ...575..493J,2010SPIE.7740E..0DJ}. In a first analysis of sectors 17 and 18, TOI-1470 was announced on 5 December 2019 as a \textit{TESS} object of interest (TOI) via the dedicated MIT \textit{TESS} data alerts public website \footnote{\url{ https://tess.mit.edu/toi-releases/}}, where a planet candidate was identified at an orbital period of 2.52713$\pm$0.00001\,d. The planet candidate was fit with limb-darkening transit models by SPOC \citep{Li:DVmodelFit2019} and successfully passed all the diagnostic tests performed on them \citep{Twicken:DVdiagnostics2018}. No further transiting planet signatures were detected in the residual light curve by the SPOC. The planet candidate, TOI-1470\,b, has an estimated radius of 2.18$\pm$1.43\,R$_{\oplus}$ and a depth of 1928$\pm$149\,ppm. With the latest release of sector 58, {\sl TESS} announced an additional transiting planet candidate with an orbital period of 36.1766$\pm$0.0002\,d, estimated radius of 2.44$\pm$1.75\,R$_{\oplus}$, and photometric depth of 2452$\pm$312\,ppm. However, as explained in Section \ref{subsec:TESS_light_curve}, this candidate was already seen in sectors 17, 18, and 24 with an orbital period twice as short as announced by \textit{TESS}. We named this candidate TOI-1470\,c.

\begin{figure*}[]
        \centering
        \includegraphics[width=0.24\textwidth]{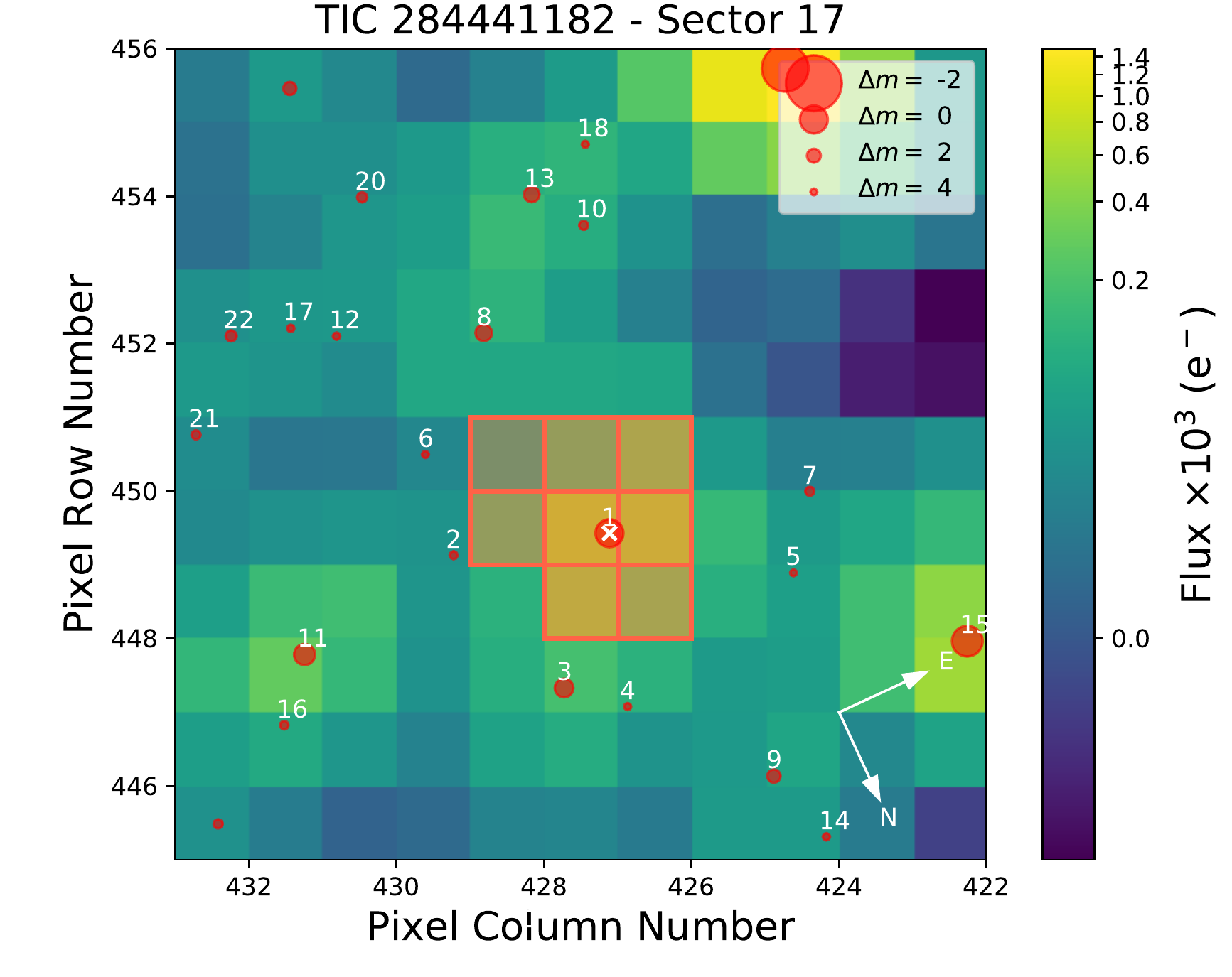}
        \includegraphics[width=0.24\textwidth]{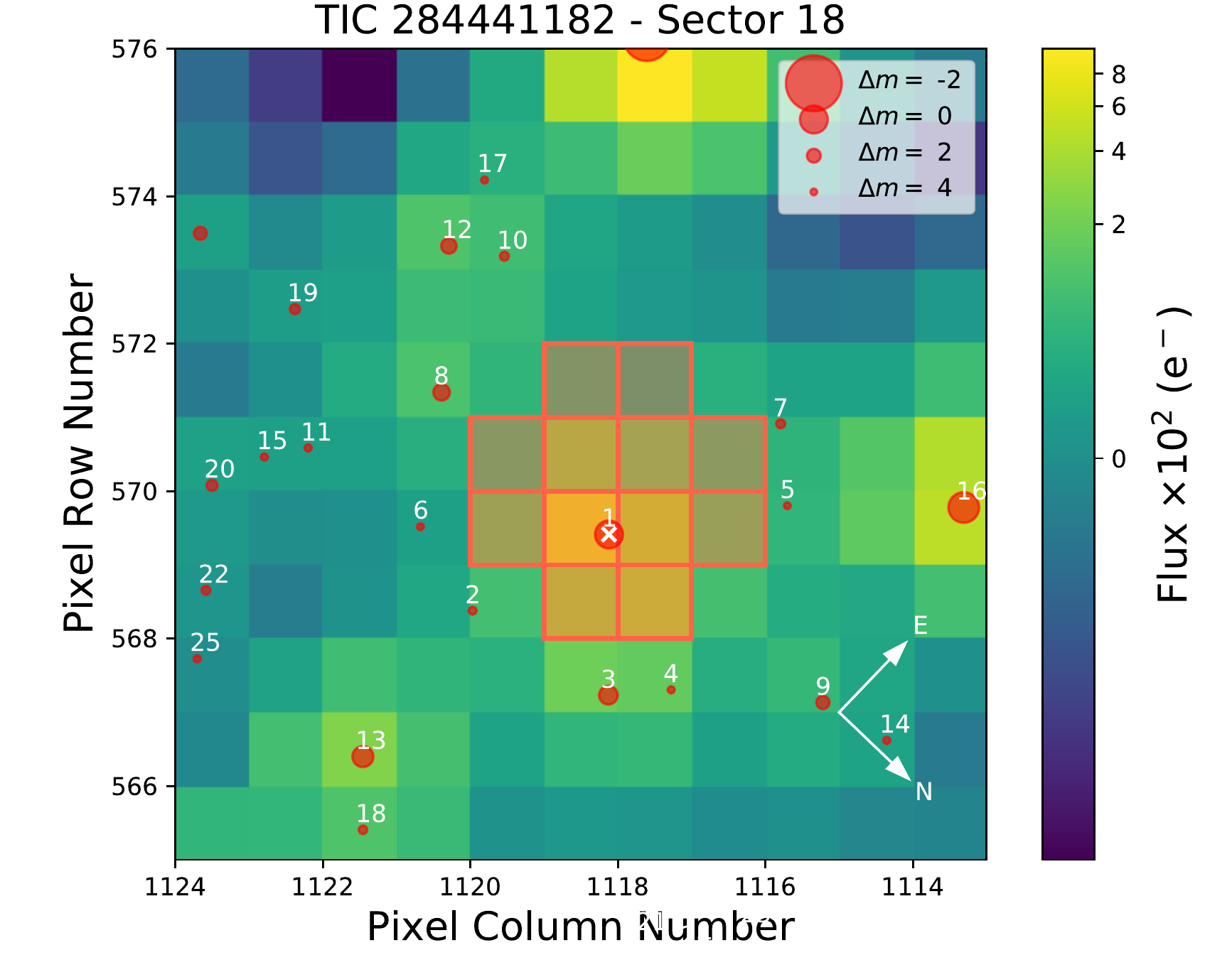}
        \includegraphics[width=0.24\textwidth]{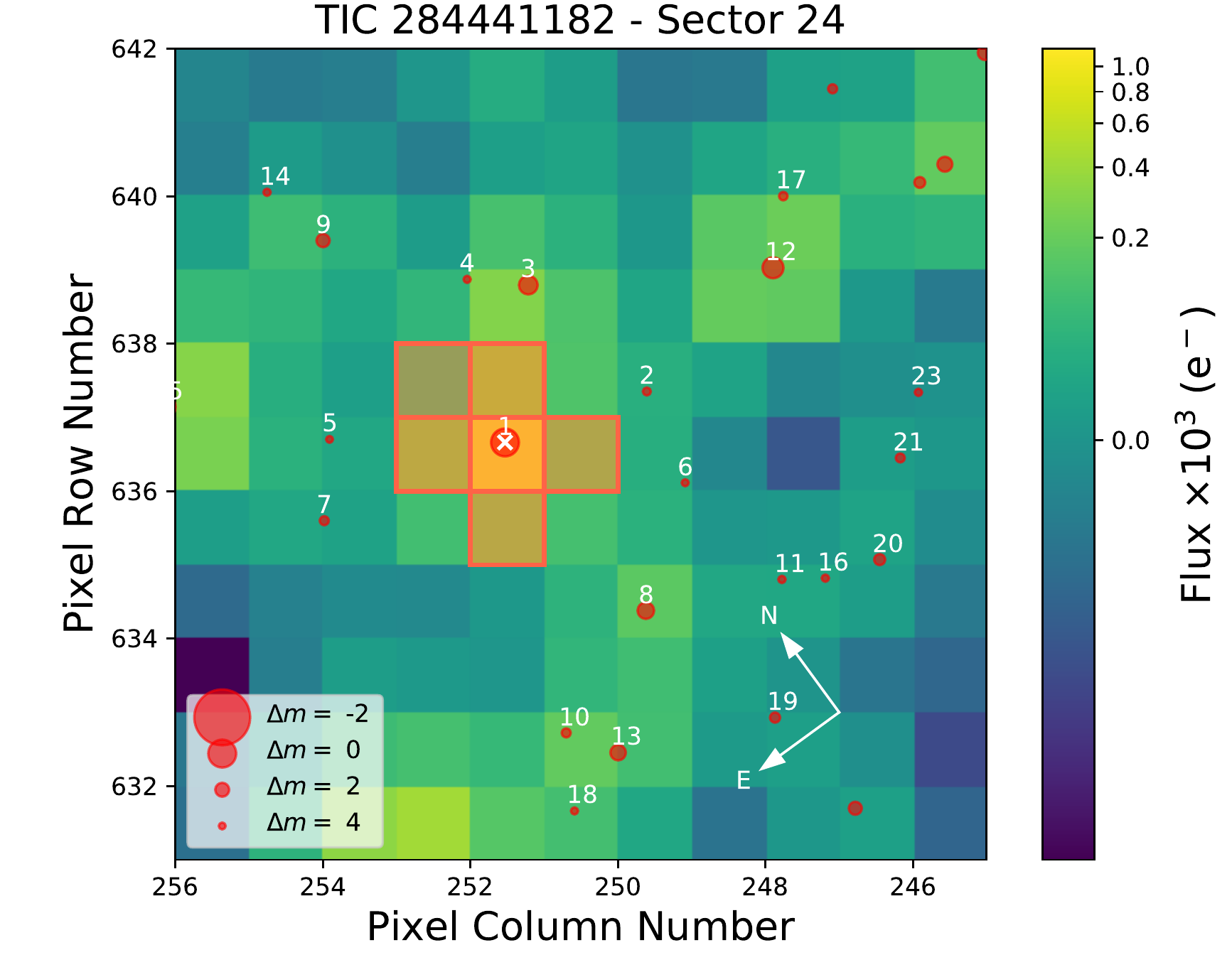}
    \includegraphics[width=0.24\textwidth]{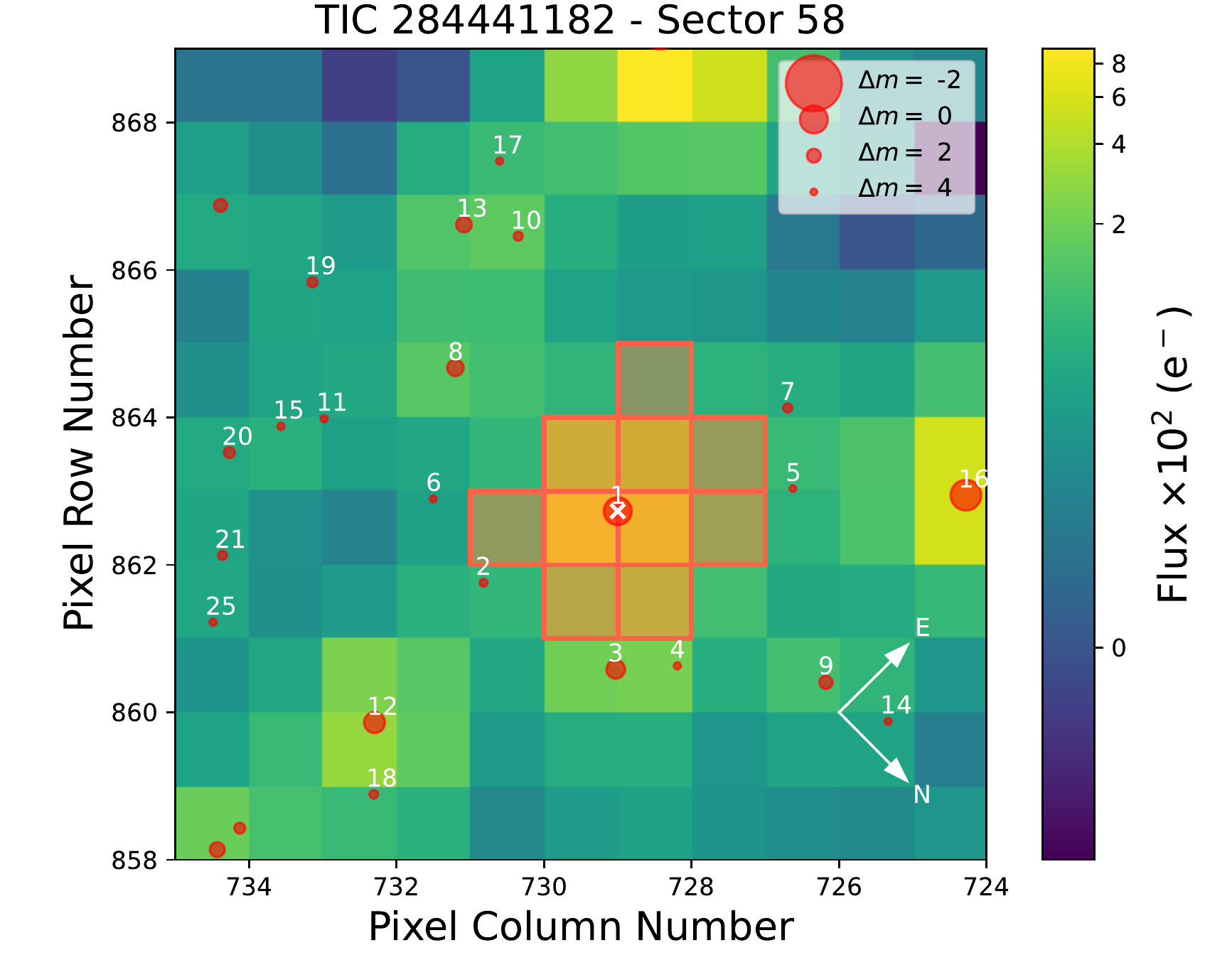}
        \caption{TPFs of TOI-1470 (cross) in \textit{TESS} sectors 17, 18, 24, and 58 (left to right). The electron counts are color-coded. The \textit{TESS} optimal photometric aperture
                per sector used to obtain the SAP fluxes is marked with red squares. The \textit{Gaia} DR2 objects with $G$-band magnitudes down to 4\,mag fainter than TOI-1470 are labeled with numbers (TOI-1470 corresponds to number 1), and their scaled brightness based on \textit{Gaia} magnitudes is shown by red circles of different sizes (inset). The pixel scale is 21\,arcsec\,pixel$^{-1}$.}
        \label{fig:apertures}
\end{figure*}

\begin{figure*}[] 
        \centering
        \includegraphics[width=\textwidth]{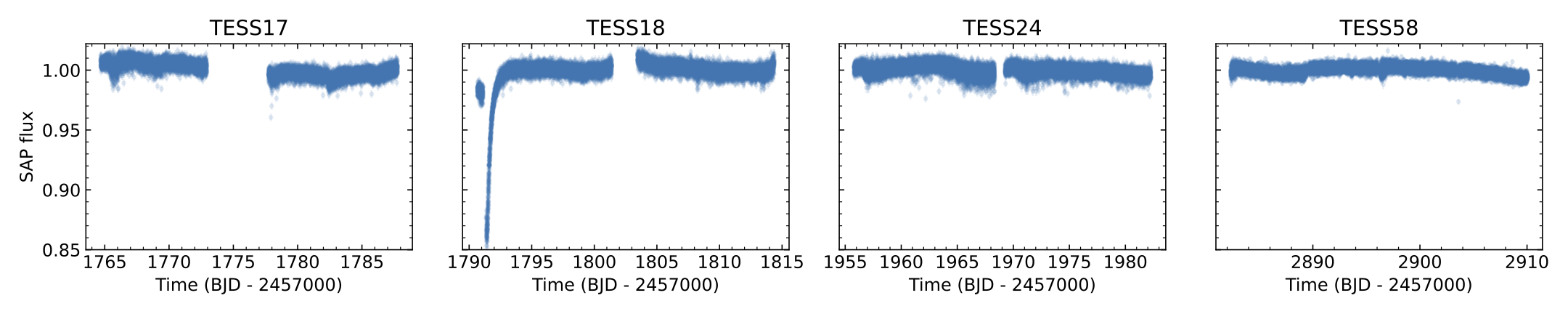}
        \includegraphics[width=\textwidth]{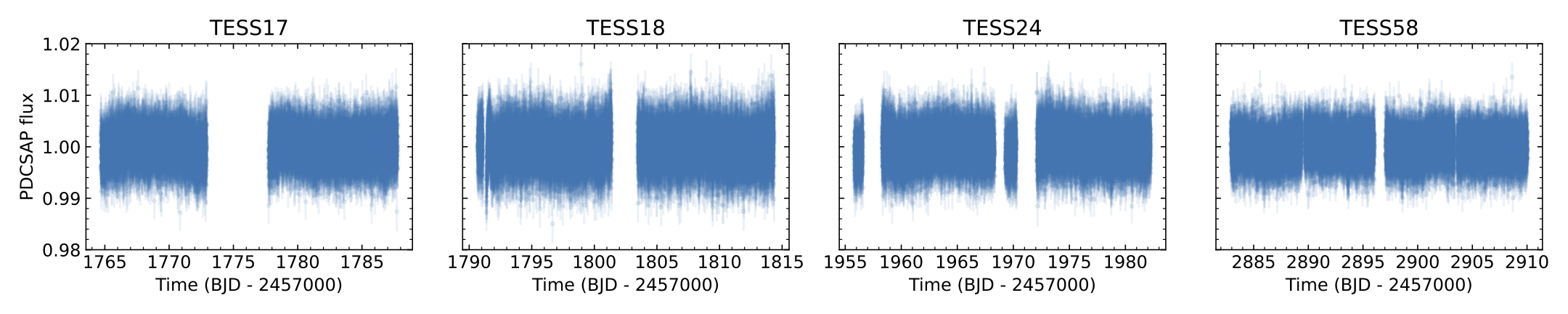}
        \includegraphics[width=\textwidth]{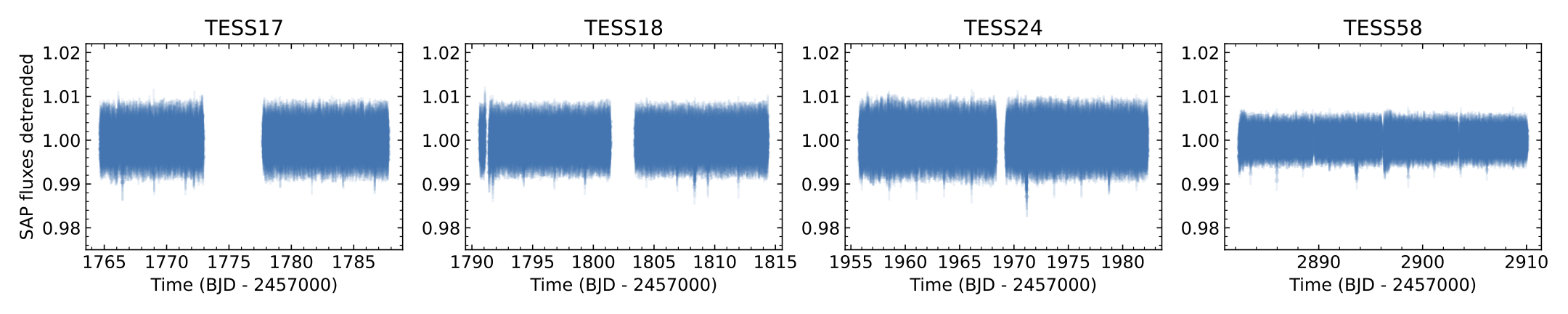}
        \caption{\textit{TESS} LCs. {\sl Upper and middle panels:} SAP and PDCSAP light curves (blue dots) of TOI-1470 as obtained by \textit{TESS} in sectors 17, 18, 24, and 58, and processed by SPOC (left to right). {\sl Lower panels:} SAP fluxes detrended and cleaned from outliers by us (blue dots) and used in our final planetary analysis. The transit features have been masked for the detrending.}
        \label{fig:SAP_and_PDCSAP}
\end{figure*}

The light curves and TPFs files for the different sectors were downloaded from the Mikulski Archive for Space Telescopes, which is a NASA-funded project. Our first step was to verify that the SAP and PDCSAP fluxes automatically computed by the pipeline are useful for scientific studies by confirming that no additional bright source contaminates the aperture photometry. Figure~\ref{fig:apertures} displays the TPFs of TOI-1470 for all four sectors using the publicly available {\tt tpfplotter}\footnote{ \url{https://github.com/jlillo/tpfplotter}} code \citep{2020A&A...635A.128A}, which overplots the \textit{Gaia} Data Release 2 (DR2) catalog \citep{2018A&A...616A...1G} on the \textit{TESS} TPFs. We confirmed that there are no additional \textit{Gaia} sources within the photometric aperture around TOI-1470 that are automatically selected by the pipeline. Therefore, we considered the extracted \textit{TESS} LC to be free from contamination from nearby stars. The original SAP and PDCSAP light curves of TOI-1470 are illustrated in the top and middle panels of Fig.~\ref{fig:SAP_and_PDCSAP}, and the SAP light curves cleaned from outliers and flattened by us are illustrated in the bottom panel of Fig.~\ref{fig:SAP_and_PDCSAP}.

We use the SAP fluxes below to search for the $P_{\rm rot}$ of the star (Sect. \ref{subsec:TESS SAP flux}) and also to derive the orbital parameters of the planets in our spectrophotometric joint fit (Sect. \ref{sec:planet orbiting toi-1470}). However, we use the PDCSAP fluxes in a blind search of other possible transiting planets (Sect. \ref{subsec:TESS_light_curve}).

\subsection{MuSCAT photometry}
\label{subsec:muscat}

One full transit of TOI-1470\,b was observed on 31 July 2020 with the multicolor simultaneous camera for studying atmospheres of transiting exoplanets (MuSCAT) instrument \citep{2015JATIS...1d5001N} at the NAOJ 1.88\,m telescope located in Okayama, Japan. MuSCAT is equipped with three 1024x1024 pixel CCDs that can be controlled independently. The three CCDs operate at $g'$ (400--550\,nm), $r'$ (550--700\,nm), and $z_{s}$ (820--920\,nm) bands using Astrodon Photometrics Generation 2 Sloan filters. The field of view of MuSCAT is 6.1x6.1\,$\rm arcmin^{2}$ with a pixel scale of 0.358\,arcsec per pixel. Image calibration and aperture photometry were performed using a custom pipeline \citep{2011PASJ...63..287F}. For our target TOI-1470\,b, the transit is detected in the $r'$ and $z_{s}$ bands (the timing and depth are consistent with \textit{TESS}). However, the light curve in the $g'$ band is very noisy and is more affected by systematics; therefore, we discarded this filter in our transit fit.

\subsection{MuSCAT2 photometry}

The second transiting-planet candidate, TOI-1470\,c, was observed on 20 September 2021 with the MuSCAT2 instrument \citep{2019JATIS...5a5001N} at the 1.52\,m Carlos S\'anchez Telescope at Observatorio del Teide, Spain. MuSCAT2 is a four-channel imager that performs simultaneous photometry in the $g'$, $r'$, $i'$, and $z_{s}$ bands, and it is equipped with four 1024 $\times$ 1024 pixel CCDs. It has a field of view of 7.4 $\times$ 7.4\,$\rm arcmin^{2}$ with a pixel scale of 0.44\,arcsec per pixel. 

Data reduction and photometric analysis were carried out using a custom-built pipeline for MuSCAT2 \citep{2020A&A...633A..28P}. The pipeline provides aperture photometry for a set of comparison stars and different aperture sizes. The aperture size for the two used reference stars was 6.09\,arcsec and the aperture size for TOI-1470 was 13.92\,arcsec. After a global optimization that took the transit model and several different sources of systematics from covariates into account, the final light curves were chosen. Our observations reveal a transit feature in $i'$ and $z_{s}$, with compatible parameters. However, the $g'$ band is more affected by systematics, and the transit feature is not visible by eye. Our target is an M star, and the flux is very low in the blue bands; therefore, we decided to exclude the $g'$ band information from our analysis. 

\subsection{MuSCAT3 photometry}

A transit of TOI-1470\,c was also observed with MuSCAT3 on 26 October 2021. MuSCAT3 is a four-channel ($g'$, $r'$, $i'$, and $z_{s}$) simultaneous imager on the LCOGT two-meter telescope (FTN) at Haleakala Observatory \citep{2020SPIE11447E..5KN} with four independent 2048 $\times$ 2048 pixel CCDs, each with a field of view of 9.1 $\times$ 9.1\,$\rm arcmin^{2}$ with a pixel scale of 0.27\,arcsec per pixel. Image calibration was performed using the {\tt BANZAI} pipeline \citep{2018SPIE10707E..0KM}, and aperture photometry was performed using the same pipeline as was used for the MuSCAT data (Section \ref{subsec:muscat}). The MuSCAT3 photometric data of TOI-1470 in the four observed channels were used in our analysis.

\subsection{LCOGT photometry}

We observed an egress of TOI-1470\,c from the Las Cumbres Observatory Global Telescope \citep[LCOGT,][]{2013PASP..125.1031B} 1.0\,m network node at Tenerife, between 1 September 2021 23:48 UT and 2 September 2021 02:08 UT. The observations were conducted in $rp$ band. The images were calibrated by the standard LCOGT {\tt BANZAI} pipeline \citep{2018SPIE10707E..0KM}. We extracted photometry using circular apertures with a radius of 16 pixels, and we removed systematic trends by differential photometry with two reference stars. The choices of aperture radii and reference stars minimized the scatter in the median normalized light-curve of the target star. The photometric scatter of $0.02 \%$ per hour integration was good enough to robustly detect the transit of TOI-1470\,c if it had been fully observed. Unfortunately, the observation only sampled the transit egress with a short post-transit baseline of about 15 minutes. We fit a transit model with fixed ephemerides obtained from the \textit{TESS} and RV data, and limb-darkening coefficients. We found that this model was marginally favored over a constant line, based on the Akaike information criterion ($\Delta \mathrm{AIC}$ = --4). We decided not to include this dataset in the final fit.

\subsection{ASAS-SN photometry}

\begin{figure}[]
        \centering
        \includegraphics[width=\columnwidth]{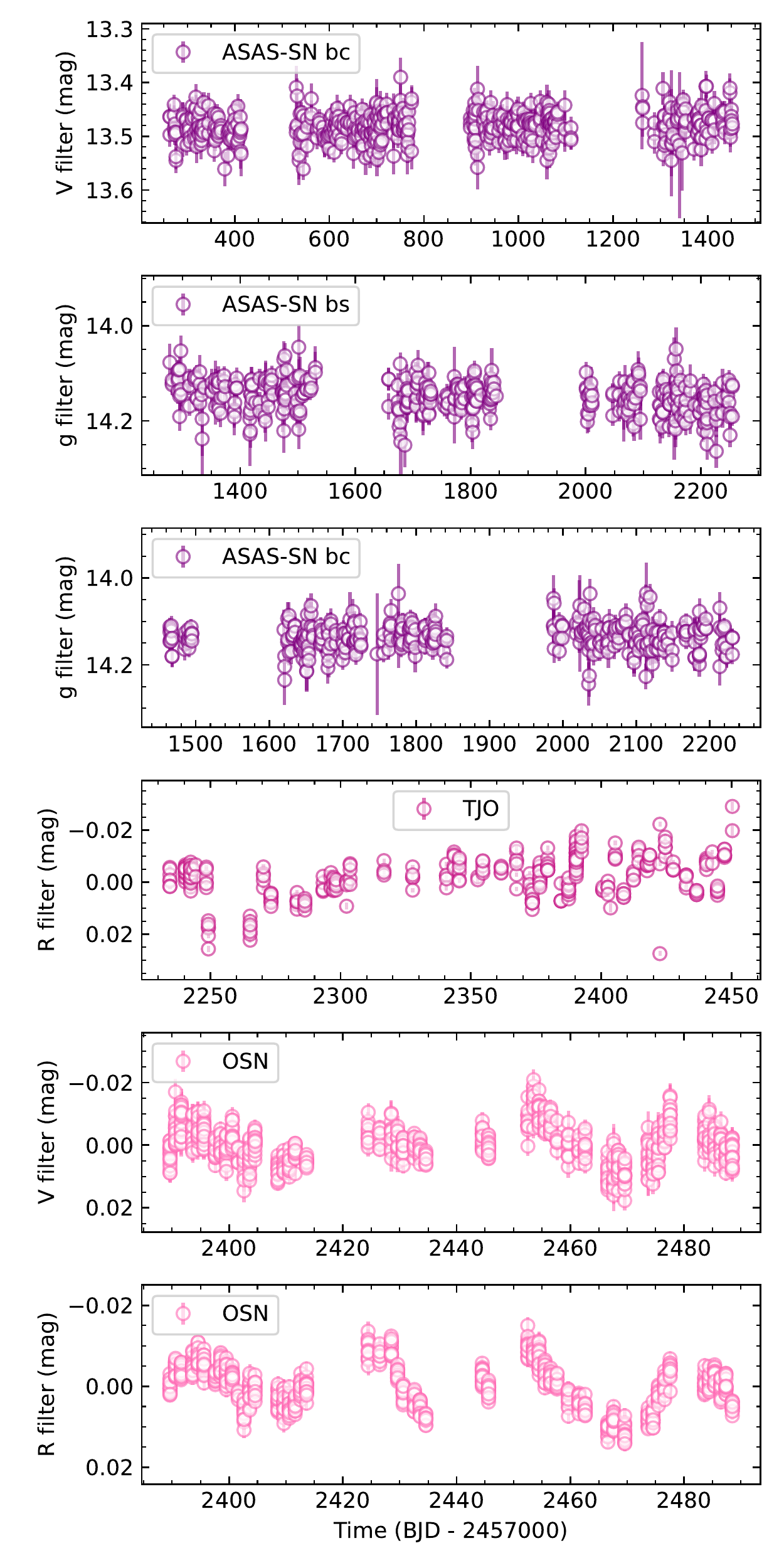}
        \caption{Photometric time series of TOI-1470 using ASAS-SN (aperture photometry), TJO (differential photometry), and OSN (differential photometry). }
        \label{fig:ASAS-SN_phot}
\end{figure}

TOI-1470 was photometrically observed by the All-Sky Automated Survey for Supernovae (ASAS-SN) project \citep{2014ApJ...788...48S, 2017PASP..129j4502K}. ASAS-SN currently consists of 24 telescopes that are distributed around the globe. These telescopes are used to survey the entire visible sky every night down to 18\,mag using robotic telescopes with a diameter of 14\,cm; the large number of telescopes minimizes the impact of poor weather. The ASAS-SN data of TOI-1470 were taken  with two different passbands between August 2015 and November 2018 ($V$ band) and between April 2018 and February 2021 ($g'$ band). The ASAS-SN aperture photometry is calibrated using the AAVSO Photometric All-Sky Survey catalog \citep[APASS,][]{2015AAS...22533616H}. The ASAS-SN detectors have a pixel scale of $8\arcsec$ projected onto the sky; the images of the stars have a typical full width at half maximum of about $15\arcsec$, so there could be some stellar blending particularly for crowded fields. We downloaded the ASAS-SN $V$- and $g'$-band photometry of TOI-1470. The different cameras (bs and bc for the $g'$ band) were treated separately in order to minimize any possible systematics. The $V$-band analysis was carried out only with the bc camera. The original photometry contained several outliers, and we applied a 2.5$\sigma$ clipping algorithm to clean the various ASAS-SN light curves. The root mean square ($rms$) of the data is $\sim$0.03\,mag. The $V$- and $g'$-band photometric time series are shown in the three top panels of Fig. \ref{fig:ASAS-SN_phot}.

\subsection{TJO photometry}

We observed TOI-1470 with the 0.8\,m Telescopi Joan Or\'o \citep[TJO,][]{2010SPIE.7740E..3KC} at the Observatori del Montsec in Lleida, Spain, from 19 January 2021 to 23 August 2021. A total of 349 images were obtained with the Johnson $R$ filter using the LAIA imager, a 4k$\times$4k CCD with a field of view of 30\,arcmin, and a scale of 0.4$\arcsec$\,$\rm pixel^{-1}$. Raw frames were corrected for dark current and bias and were flat-fielded using the ICAT pipeline \citep{2006IAUSS...6E..11C} of the TJO. The aperture photometry of TOI-1470 was extracted with the {\tt AstroImageJ} software \citep{2017AJ....153...77C} by using an optimal aperture size that minimized the $rms$ of the resulting relative fluxes. To derive the differential photometry of TOI-1470, we selected the 12 brightest comparison stars in the field that did not show any variability. Then, we employed our own pipelines to remove outliers and measurements affected by poor observing conditions or a low signal-to-noise ratio. The $rms$ of the TJO differential photometry after the removal of outliers is 7\,mmag. The TJO $R$-band light curve is presented in the fourth panel of Fig.~\ref{fig:ASAS-SN_phot}.

\subsection{OSN photometry}

We also monitored TOI-1470 using a 150 cm Ritchey-Chr\'etien telescope at the Observatorio de Sierra Nevada (OSN), Spain. The T150 telescope is equipped with an Andor Ikon-L 2k$\times$2k CCD camera, which delivers images with a field of view of 7.92\arcmin$\times$7.92{\arcmin} \citep{2022A&A...663A..48Q}. The observations were collected with both Johnson $V$ and $R$ filters at a total of 54 observing epochs over 99 days in the period June--September 2021. Per observing epoch, we typically obtained 20 individual measurements per filter, with an integration time of 70\,s and 50\,s for $V$ and $R$ filters, respectively. We obtained aperture photometry from the unbinned frames, which were bias subtracted and properly flat-fielded beforehand. We explored different aperture sizes and the circular aperture with radius of 24 pixels produced the lowest standard deviation photometric light curve for both filters. From the final OSN photometric time series, we removed low-quality data obtained under poor observing conditions or at very high air masses. The final OSN light curves have an $rms$ of 6\,mmag and 5\,mmag for the $V$ and $R$ filters, respectively. The $V$- and $R$-band OSN photometric time series are shown in the bottom panels of Fig.~\ref{fig:ASAS-SN_phot}.

\subsection{Palomar 5m AO image}

\begin{figure}
        \centering
        \includegraphics[width=\columnwidth]{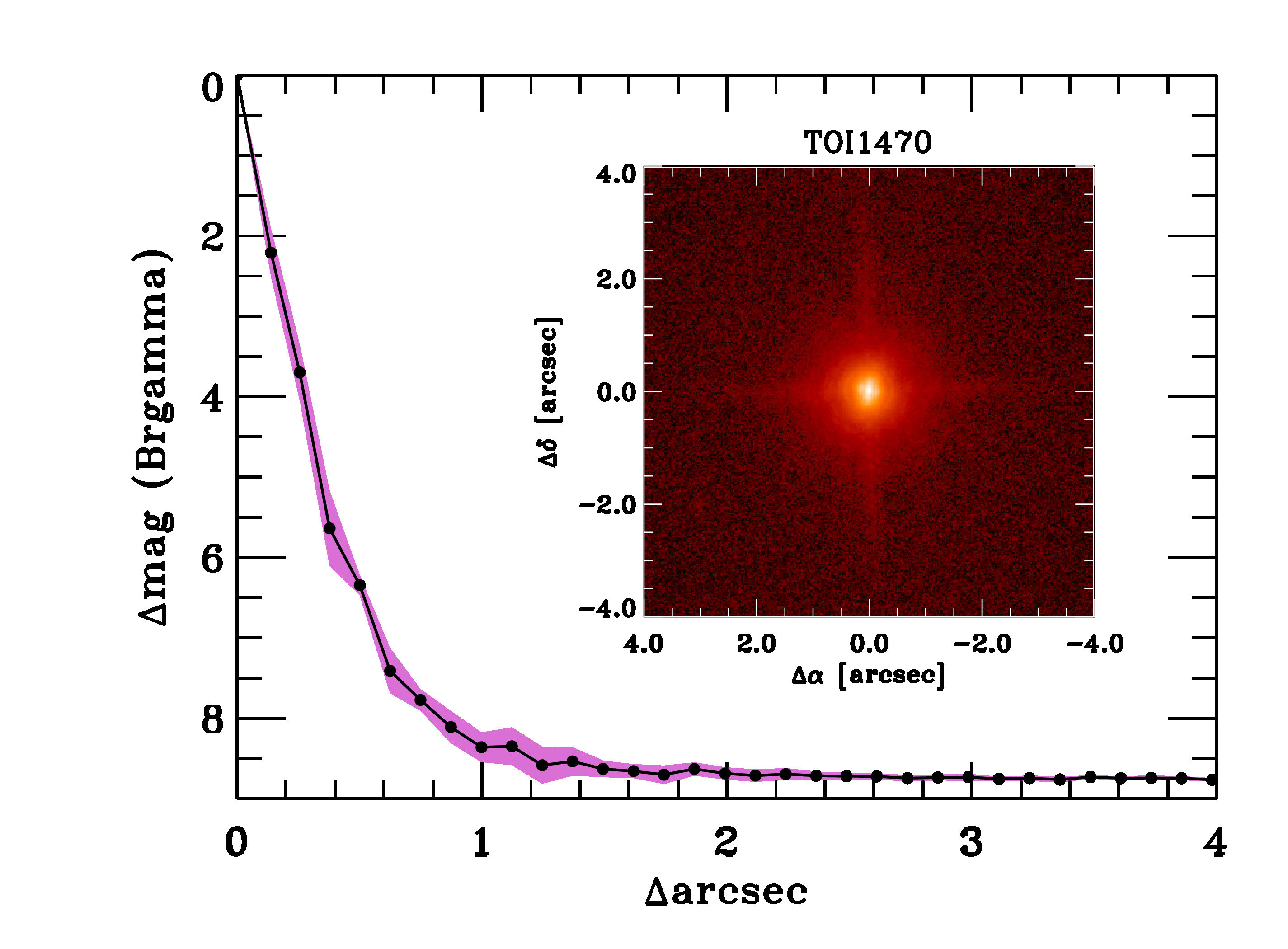}
        \caption{Relative $5\sigma$ detection limit magnitude as a function of the angular separation for the Palomar AO using the Br-$\gamma$ filter (dotted black line). The inset shows the reconstructed PHARO AO image.}
        \label{fig:Palomar_image}
\end{figure}

TOI-1470 was observed on 8 January 2020 UT using the PHARO near-infrared camera \citep{2001PASP..113..105H} for the adaptive optics (AO) system on the Palomar 5\,m telescope. The observations were carried out in the narrow-band Br-$\gamma$ filter using a CCD that has a pixel scale of 0.025$\arcsec$\,$\rm pixel^{-1}$. The image of the star has a full width at half maximum of  0.12$\arcsec$. PHARO provides AO imaging  with output data products including a reconstructed image with robust contrast limits on companion detections \citep[e.g.,][]{2016ApJ...829L...2H}. Figure~\ref{fig:Palomar_image} shows our final $5\sigma$ contrast curves and the Br-$\gamma$ filter reconstructed AO image. In the PHARO AO image, TOI-1470 has no stellar companion with a contrast of $\Delta m \approx$ 7\,mag at projected angular separations in the interval 0.5--4.0$\arcsec$ (i.e., 25--208\,au at the distance of the system).

\subsection{CARMENES spectroscopic data}
\label{subsec:carmenes_spectroscopy}

After the announcement of TOI-1470 as a \textit{TESS} object of interest, TOI-1470 was spectroscopically monitored from 3 June 2020 to 17 January 2021. We obtained a total of 44 high-resolution spectra using the fiber-fed échelle spectrograph CARMENES. CARMENES is installed at the 3.5\,m telescope of the Calar Alto Observatory in Almer\'ia (Spain). It was specifically designed to deliver high-resolution spectra at optical (resolving power $\mathcal{R} \approx$ 94,600) and near-infrared ($\mathcal{R} \approx$ 80,500) wavelengths from 520 to 1710\,nm. CARMENES has two different channels, one channel for the optical (the VIS channel) and the other for the near-infrared (the NIR channel), with a break at 960\,nm \citep{2016SPIE.9908E..12Q}. All data were acquired with integration times of 1800\,s (which is the maximum exposure employed by us for precise RV measurements) and followed the data flow of the CARMENES GTO program \citep{2023A&A...670A.139R}. CARMENES raw data are automatically reduced with the {\tt caracal} pipeline \citep{2016SPIE.9910E..0EC}. Relative RVs are extracted separately for each \'echelle order using the {\tt serval} software \citep{2018A&A...609A..12Z}. The final VIS and NIR RVs per epoch are computed as the weighted RV mean over all \'echelle orders of the respective spectrograph.

\begin{figure}[]
        \centering
        \includegraphics[width=\columnwidth]{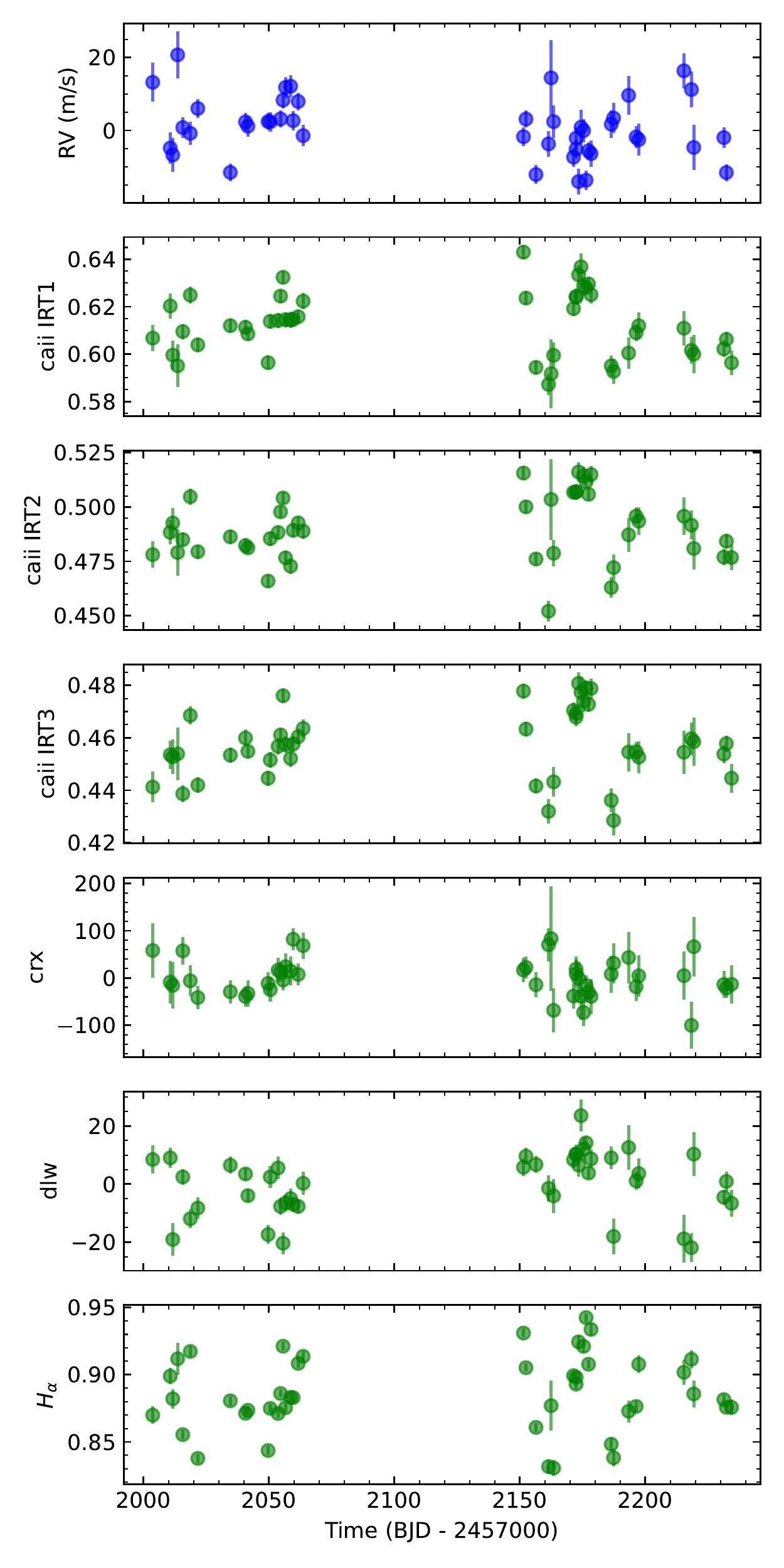}
        \caption{CARMENES VIS RV time series (blue dots) and various spectroscopic activity indicators (green dots) of TOI-1470. }
        \label{fig:toi1470_activity_time}
\end{figure}

Our first step was to search for RV outliers in the CARMENES time series and for data points with very large error bars (more than three times the mean error bar size, 3.5\,m\,s$^{-1}$). However, no RV data points were discarded, and the final CARMENES dataset has 44 RV measurements with an $rms$ of 8.1\,m\,s$^{-1}$; it is displayed in the top panel of Fig.~\ref{fig:toi1470_activity_time}. The individual CARMENES relative RVs and their uncertainties are listed in Table~\ref{tab:toi1470_rv_act_data}.

At high spectral resolution, the profile of the stellar lines may change due to photospheric and chromospheric activity, which has an impact on accurate RV measurements; it is crucial to disentangle the effects of stellar activity from the Keplerian signals. The CARMENES {\tt serval} pipeline provides measurements for a number of spectral features that are considered indicators of stellar activity, such as the differential line width (dLW), H$\alpha$,  the Ca\,{\sc ii} infrared triplet $\lambda\lambda$8498, 8542, 8662\,\AA~(IRT), and the chromatic index (CRX). The latter determines the RV--$\log \lambda$ correlation, and it is used as an indicator of the presence of stellar active regions. All these indices may have a chromospheric component in active M dwarfs and are defined by \cite{2018A&A...609A..12Z}. Additionally, as part of the data processing, we calculated measurements of molecular absorption bands of two species: TiO and VO, together with pseudo-equivalent widths (pEW) for different indices after the subtraction of a reference star spectrum as described by \cite{Schofer2019}. The CARMENES activity indicators of TOI-1470 are shown in Fig.~\ref{fig:toi1470_activity_time}, where data points deviating significantly (more than 2.5 $\sigma$) from the sequence or data with very large error bars have been removed from the time series. We used these activity indices to analyze the properties of the parent star.

\section{The star TOI-1470}
\label{sec:TOI-1470}

\subsection{Stellar properties}

 \begin{table}[!t]
        \centering
        \small
        \caption{Stellar parameters of TOI-1470.} 
        \label{tab:stellar_properties_TOI-1470}
        \begin{tabular}{lcr}
                \hline
                \hline
                \noalign{\smallskip}
                Parameter & Value & Reference \\ 
                \noalign{\smallskip}
                \hline
                \noalign{\smallskip}
                \multicolumn{3}{c}{\em Basic identifiers and data}\\
                \noalign{\smallskip}
                TIC                                                                     & 284441182 & Stas18 \\
                Karmn                                                           & J00403+612 & Cab16\\
                2MASS                                                   & J00402129+6112490  & 2MASS\\
                Sp. type                            & M1.5\,V      & This work \\
                $V$ (mag) & $13.454 \pm 0.020$ & Cifu20$^a$\\
                $G$ (mag) & $12.5425     \pm 0.0003$ &  \textit{Gaia} DR3$^a$\\
                $T$ (mag) & $11.475 \pm 0.007$ & TIC$^a$\\ 
                $J$ (mag) & $10.154      \pm 0.020$ &  2MASS$^a$\\
                \noalign{\smallskip}
                \multicolumn{3}{c}{\em Astrometry and kinematics}\\
                \noalign{\smallskip}
                $\rm \alpha$ (J2000) &   00:40:21.39 &  \textit{Gaia} DR3\\
                $\rm \delta$ (J2000) &   +61:12:48.2  &  \textit{Gaia} DR3\\    
                $\mu _{\alpha} \cos \delta$ ($\rm mas\,yr^{-1}$) & $48.576 \pm 0.013$ & \textit{Gaia} DR3\\
                $\mu _{\delta}$ ($\rm mas\,yr^{-1}$) & $-55.734 \pm 0.013$ & \textit{Gaia} DR3\\
                $\varpi$ (mas) & $19.3277        \pm 0.0131$ &  \textit{Gaia} DR3 \\
                $d$ (pc) & $\rm 51.9503 \pm 0.075$  &  \\
                $\gamma$ (km s$^{-1}$)                  & $\rm 3.54     \pm0.50$ & \textit{Gaia} DR3\\
                $U$ (km s$^{-1}$)  & $-10.97 \pm 0.42$ & This work \\
                $V$ (km s$^{-1}$)   & $-3.84 \pm 0.69$ & This work \\
                $W$ (km s$^{-1}$)   & $-14.31 \pm 0.03$ & This work \\
                Galactic population                 & Young disk & This work \\
                \noalign{\smallskip}
                \multicolumn{3}{c}{\em Fundamental parameters and abundances}\\
                \noalign{\smallskip}
                $T_{\rm eff}$ (K)  & $\rm 3709  \pm 11 $ & Marf21 \\
                $\log g$ (cgs) & $\rm 5.09 \pm 0.09$ & Marf21\\
        $\log g$ (cgs)  &   4.77 $\pm$ 0.02  &    This work$^b$ \\
                $\rm [Fe/H]$ (dex) & $\rm -0.00\pm 0.03$ & Marf21\\
                $\rm [Mg/H]$ (dex) & $\rm +0.07\pm 0.11$ & Tabe\\
                $\rm [Si/H]$ (dex) & $\rm -0.13\pm 0.09$ & Tabe\\
                $L$ $(\rm L_{\odot})$ & $\rm 0.0376 \pm 0.0001$ & This work\\
                $R$ ($\rm R_{\odot}$) & $\rm 0.469 \pm 0.003$ & This work\\
                $M$ ($\rm M_{\odot}$) & $\rm 0.471\pm 0.011$ & This work\\
        $v \sin i_\star$ ($\mathrm{km\,s^{-1}}$) & $<2.0$       &  Marf21\\
                \noalign{\smallskip}
                \multicolumn{3}{c}{\em Activity and age}\\
                \noalign{\smallskip}
                $P_{\rm rot}$ (d) & $29 \pm 3 $ & This work\\
                $\log{L_{\rm X}}$ (erg\,s$^{-1}$)   & $\sim$ 27.9 & This work \\
                Age (Gyr) & 0.6--2.0 & This work \\
                \noalign{\smallskip}
                \hline
        \end{tabular}
        \tablebib{
                2MASS: \cite{2006AJ....131.1163S};
                Cab16: \cite{2016csss.confE.148C}; 
                Cifu20: \cite{2020A&A...642A.115C};
                \textit{Gaia} DR3: \cite{2021A&A...649A...1G};
                Marf21: \cite{2021A&A...656A.162M};
                Stas18: \cite{2018AJ....156..102S};  
                Tabe: Tabernero et al. in prep.; 
                TIC: \cite{2021arXiv210804778P}.
        }
        \tablefoot{
                \tablefoottext{a}{See \citet{2020A&A...642A.115C} for multiband photometry different from {\em Gaia} $G$ and 2MASS $J$.}
                \tablefoottext{b}{Computed from $M$ and $R$.}
        }
 \end{table}

\begin{figure}[] 
        \centering
        \includegraphics[width=\columnwidth]{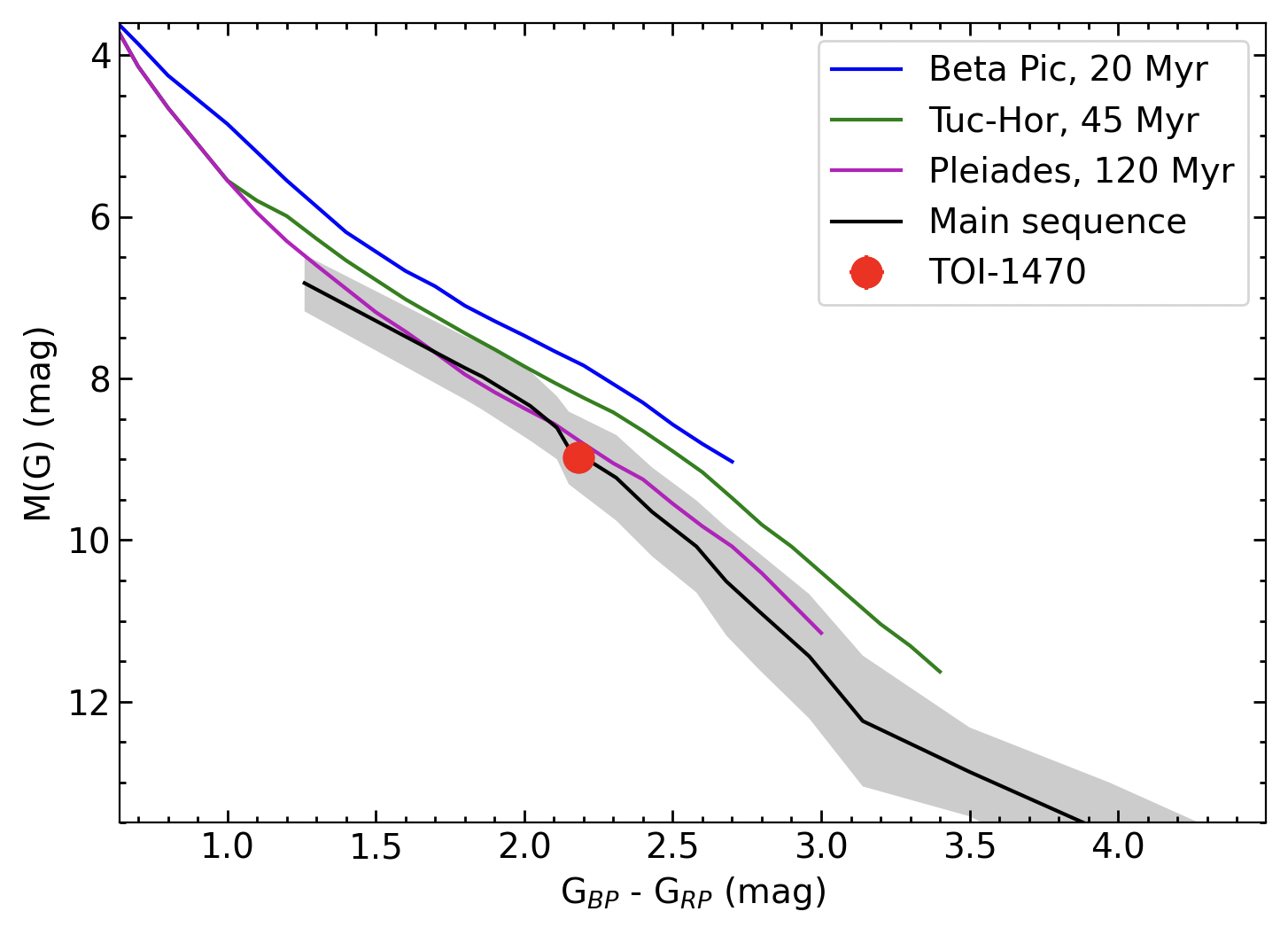}
        \caption{Location of TOI-1470 (red circle) in the \textit{Gaia} color-magnitude diagram. The young stellar sequences of the $\beta$~Pic (blue) and Tucana-Horologium (green) moving groups and the Pleiades star cluster (purple) are taken from \citet{2018AJ....156..271L}, and the main sequence of field stars (black) from \citet{2020A&A...642A.115C}. The gray area represents the dispersion observed among stars on the main sequence.}
        \label{fig:main-sequence}
\end{figure}

TOI-1470 (2MASS J00402129+6112490) is not an especially well-studied star and does not appear very often in the literature, except in large catalogs. In Table~\ref{tab:stellar_properties_TOI-1470} we provide optical and near- and mid-infrared photometry extracted from the \textit{Gaia}, 2MASS, and AllWISE catalogs \citep{2016A&A...595A...1G, 2006AJ....131.1163S, 2010AJ....140.1868W}, together with the \textit{Gaia} DR3 trigonometric parallax, proper motion, and equatorial coordinates \citep{2021A&A...649A...1G}. TOI-1470 is located at a distance of 51.950$\pm$ 0.075\,pc from the Sun and has optical and infrared colors typical of M1.5\,V type stars, without obvious evidence of extinction at short wavelengths. The location of TOI-1470 in the \textit{Gaia} color-magnitude diagram is shown in Fig.~\ref{fig:main-sequence}. It is obvious that this star is not overluminous (i.e., it is not young) and has absolute magnitudes compatible with a main-sequence M1--2 star. All stellar sequences shown in the figure were built using \textit{Gaia} photometry and parallaxes (see \citealt{2018AJ....156..271L} for the young isochrones and \citealt{2020A&A...642A.115C} for the main sequence).

The stellar atmospheric parameters ($T_{\rm eff}$, $\log{g}$, and [Fe/H]) of TOI-1470 from CARMENES VIS and NIR spectra have been recently analyzed by \cite{2021A&A...656A.162M} by means of the {\sc SteParSyn} code\footnote{\url{https://github.com/hmtabernero/SteParSyn/}} \citep{tab22}. The resulting values (listed in Table~\ref{tab:stellar_properties_TOI-1470}) are $T_{\rm eff}$ = 3709$\pm$11\,K, $\log{g}$ = 5.01$\pm$0.09\,dex, and [Fe/H] = $-$0.00$\pm$0.03\,dex. We used the spectroscopically derived $T_{\rm eff}$, the bolometric luminosity obtained following \cite{2020A&A...642A.115C}, and the Stefan-Boltzmann law to compute the stellar radius, which is $R = 0.469 \pm 0.003$ R$_\odot$. The mass of TOI-1470 was then computed after the mass-radius relation of \cite{2019A&A...625A..68S} based on eclipsing binary stars. We obtained $M = 0.471 \pm 0.011$\,M$_\odot$. The stellar mass and radius are given in Table~\ref{tab:stellar_properties_TOI-1470}, together with the surface gravity directly derived from these parameters: log\,$g =  4.77 \pm 0.02$ (cgs), which differs from the spectroscopic value at the 3\,$\sigma$ level. From the spectral fitting, we also tabulate an upper limit on the projected rotational velocity of the star, $v$\,sin\,$i \leq$ 2\,km\,s$^{-1}$. Finally, we employed the spectral synthesis method, together with the PHOENIX BT-Settl atmospheric models \citep{all12} and the radiative transfer code {\tt Turbospectrum} \citep{ple12}, to determine Mg and Si abundances of TOI-1470. We measured [Mg/H] = +0.07 $\pm$ 0.11\,dex and [Si/H] = --0.13 $\pm$ 0.09\,dex. Further details of the procedure followed will be provided by Tabernero et al. (in prep.).

The Galactic space velocities $U V W$ of TOI-1470 were derived using the \textit{Gaia} coordinates, proper motion, and RV with the formulation developed by \citet{1987AJ.....93..864J}. The $U V W$ components in the directions of the Galactic center, Galactic rotation, and north Galactic pole, respectively, are given in Table~\ref{tab:stellar_properties_TOI-1470}. The right-handed system was used, and we did not subtract the solar motion from our calculations. The uncertainties associated with each space velocity component were obtained from the observational quantities and their error bars. TOI-1470 has kinematics typical of a young disk, indicating an age $<$ 2\,Gyr. However, gyrochronologically, the stellar rotation period is very similar to that of the Praesepe cluster \citep[600--800\,Myr,][]{2018ApJ...863...67G} for stars of similar mass. Therefore, we adopted an age in the range 0.6--2.0\,Gyr for TOI-1470.


\begin{figure*}[]
        \centering
        \includegraphics[width=\textwidth]{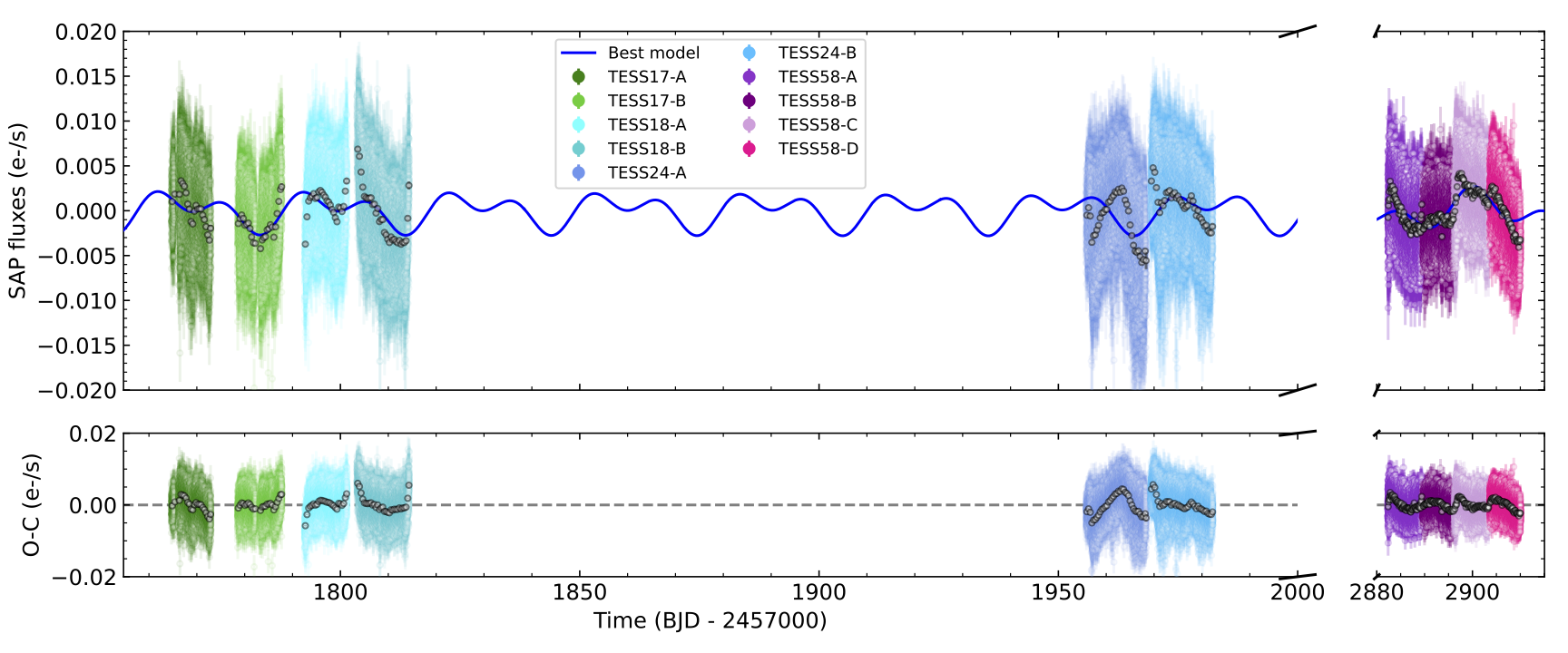}
        \caption{TOI-1470 \textit{TESS} SAP photometric fluxes for the four sectors. {\em Top panel}: Each half sector is shown with a different color normalized to a common reference by fitting two sinusoidal functions. The best model fit is plotted as the solid blue line, and the binned photometric data are indicated with gray dots. 
    {\em Bottom panel}: Residuals as a function of time.}
        \label{fig:SAP-flux_two_sin_mod}
\end{figure*}

\begin{figure}[]
        \centering
        \includegraphics[width=\columnwidth]{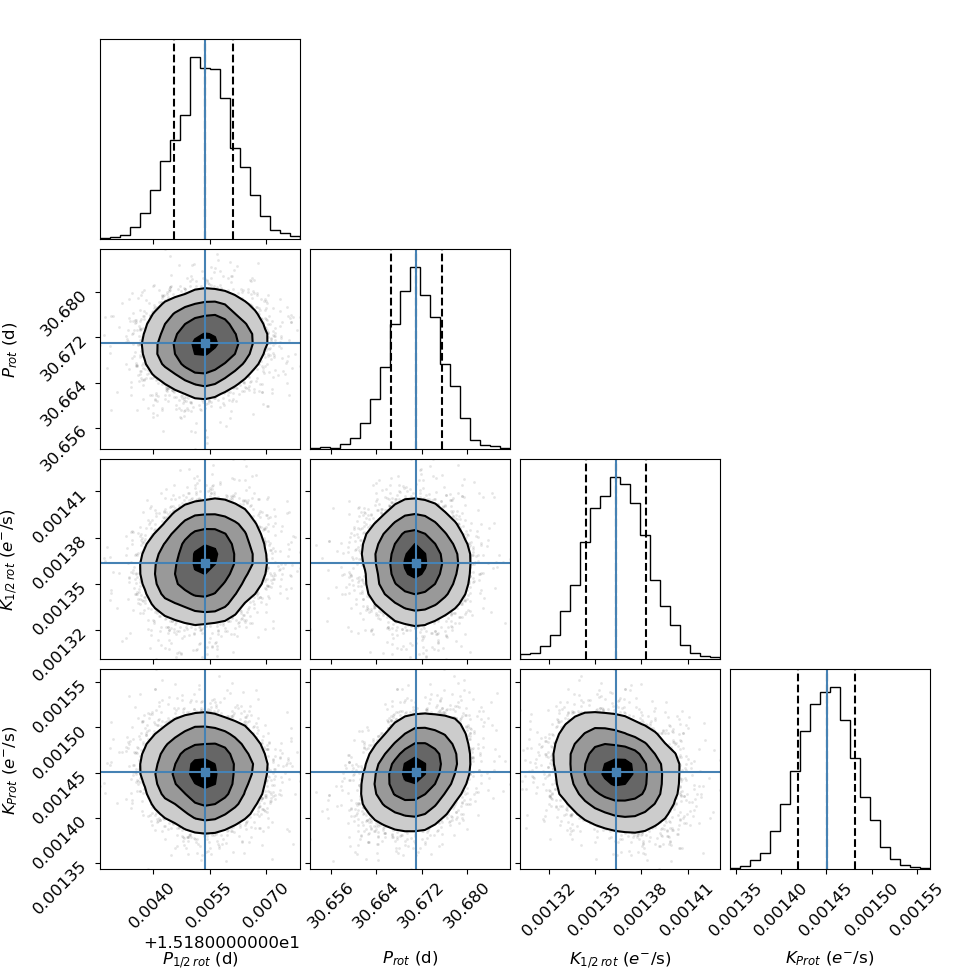}
        \caption{Posterior distributions of the sinusoidal fit to the \textit{TESS} SAP fluxes. The vertical dashed lines correspond to the 16\%~and 84\%~quantiles (1 $\sigma$ uncertainty). The solid blue lines stand for the median values of the distributions.}
        
        \label{fig:SAP-corner-plot}
\end{figure}

\subsection{Stellar variability and rotation period}
\label{sec:analysis}

M dwarfs can exhibit a wide range of stellar activity levels from inactive to very active stages, with intensities that are orders of magnitude greater than what is commonly observed in the Sun \citep{2020Sci...368.1477J}. The most active stars show inhomogeneities on their surface, such as dark starspots that corotate with the star \citep{1997A&A...327.1114L,2005ApJ...621..398O}. It is well known that stellar activity can cause line asymmetries that hinder the very precise measurement of the line center, and consequently induce an apparent RV shift, which may mimic a Keplerian signal or hide the presence of a real planet orbiting the star \citep{2015ApJ...812...42B}. Before exploring the CARMENES RVs in search for or confirming planets, we therefore analyzed all photometric and spectroscopic time series of established activity indices available to us in order to identify the characteristic frequencies of the stellar variations and, if possible, the rotation period of TOI-1470. Following previous work \citep[e.g.,][]{2014ApJS..211...24M, 2016A&A...595A..12S, 2018A&A...612A..89S, 2019A&A...624A..27G}, early-M dwarfs typically exhibit measurable rotation periods ranging from 20\,d to $\sim$100\,d and magnetic activity cycles of several hundred to one thousand days.

\subsubsection{Stellar rotation period with \textit{TESS} light curves}
\label{subsec:TESS SAP flux}

We analyzed the SAP fluxes (uncorrected for instrumental features; \citealt{twicken:PA2010SPIE,2020ksci.rept....6M}) of all three \textit{TESS} sectors. First, we cleaned the original SAP fluxes from significantly deviating points and masked the beginning of sector 18, which shows strong instrumental effects. The model assumed that each sector has a different flux offset in order to bring all sectors to a common reference. In addition, we also applied intrasector flux offsets to account for possible drifts in the photometry of the intervals before and after the \textit{TESS} data downlink time. Therefore, in practice, we divided the four \textit{TESS} light curves into ten chunks. The top panel of Fig.~\ref{fig:SAP_and_PDCSAP} shows that the SAP fluxes show no strong instrumental effects, and some low-amplitude variability becomes evident from the data. We modeled the \textit{TESS} SAP fluxes using two sinusoidal functions \citep{2011A&A...528A...4B}, where two of the free parameters are the period at the rotation period of the star ($P_{\rm rot}$) and the period at approximately half of the stellar rotation period ($P_{\rm 1/2\,rot}$). This approach can take into account the stellar variability produced by spots located at different latitudes of the stellar surface. For the main parameter, we allowed $P_{\rm rot}$ to vary between 20 and 50\,d with a uniform distribution; all other parameters (light-curve amplitudes and offsets) were explored from initial uniform distributions with a wide range of possible values. The fit was performed using the {\tt exoplanet} toolkit \citep{2021JOSS....6.3285F} for probabilistic modeling. {\tt exoplanet} extends the $PyMC3$ language to support many of the custom functions and distributions required when fitting exoplanet datasets.

We found that the most favorable period for $P_{\rm rot}$ is 30.7\,d. The flux amplitude of the variability associated with $P_{\rm 1/2\,rot}$ is about half of the amplitude of the main periodicity. Figure~\ref{fig:SAP-flux_two_sin_mod} illustrates the \textit{TESS} SAP fluxes together with the best model, and Figure~\ref{fig:SAP-corner-plot} shows the resulting corner plot of the fit (the flux offsets and jitters were excluded for clarity). We remark that all priors were uniform distributions, but the distributions of the posteriors are quite Gaussian-like. Fig.~\ref{fig:SAP-flux_two_sin_mod} shows that the best model nicely reproduces the flux variability from sector to sector, except at the beginning of the sectors. This is likely due to instrumental effects, which remain uncorrected in the SAP fluxes. We can ascribe the 30.7\,d periodicity to the rotation period of TOI-1470, although as discussed in the next sections, this value might be affected by differential rotation.

\subsubsection{ASAS-SN, TJO, and OSN light curves \label{asas-tjo-sno}}

\begin{figure}[]
        \centering
        \includegraphics[width=\columnwidth]{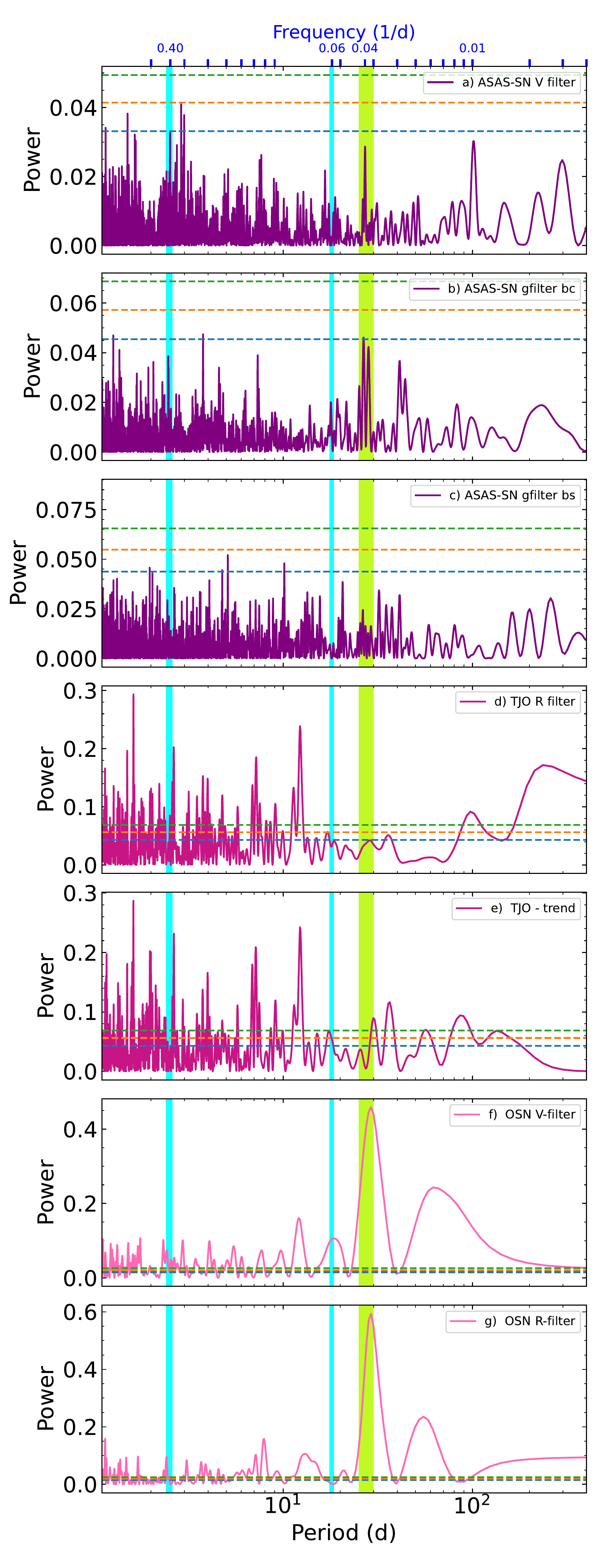}
        \caption{GLS periodograms of the ASAS-SN $V$ and $g$, TJO $R$, and OSN $V$ and $R$ light curves. In all panels, the horizontal dashed lines indicate FAP levels of 10\% (blue), 1\% (orange), and 0.1\% (green). The orbital period of the 2.5d transiting planet and another interesting signal at 18.08\,d identified in the RV data are marked with vertical blue lines. The green band indicates the region in which most of the spectroscopic activity indicators have their highest GLS peaks.}
        \label{fig:GLS_phot_multiplot}
\end{figure}

\begin{figure}[]
        \centering
        \includegraphics[width=\columnwidth]{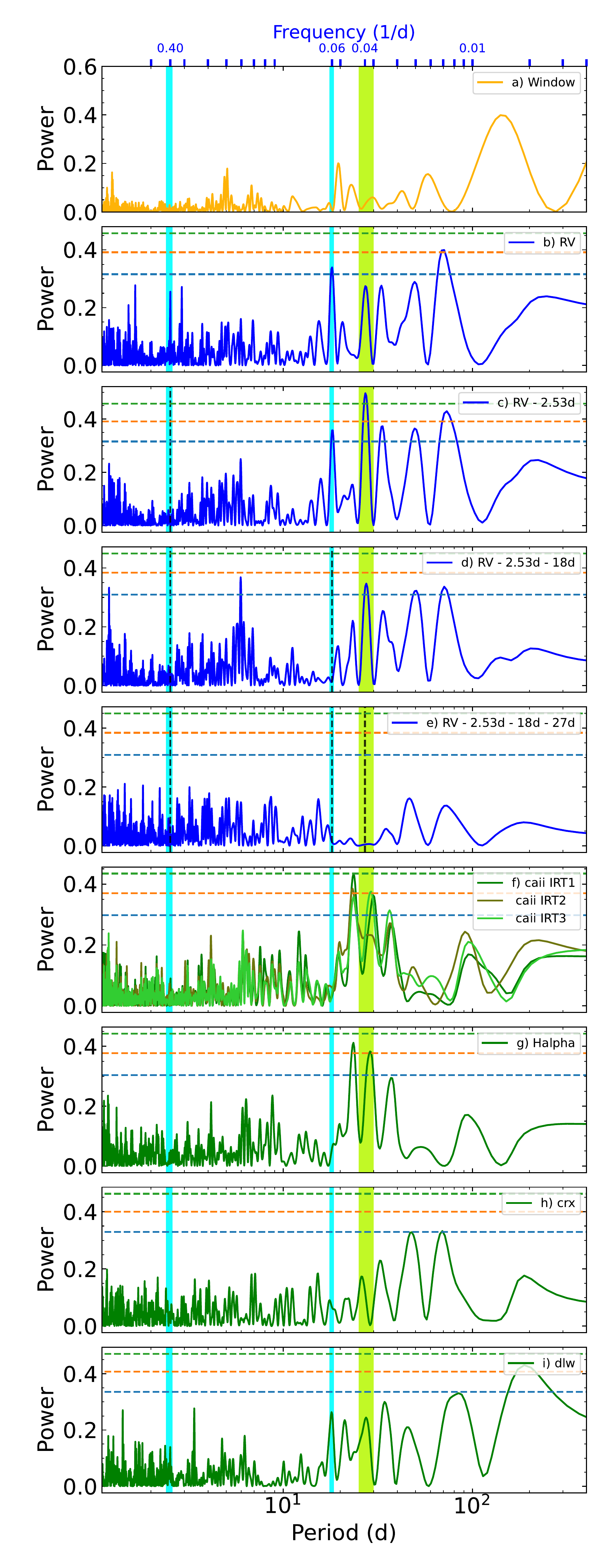}
        \caption{Window function (top panel) and GLS periodograms of CARMENES RVs (panels b--e) and various spectroscopic indices (panels f--i) between 1.1 and 350\,d. The horizontal and vertical lines and bands are the same as in Figure~\ref{fig:GLS_phot_multiplot}. }
        \label{fig:GLS_activity_multiplot}
\end{figure}

We computed the generalized Lomb-Scargle (GLS) periodograms \citep{2009A&A...496..577Z} for all ASAS-SN, TJO, and OSN light curves. The periodograms are shown in Fig.~\ref{fig:GLS_phot_multiplot}, where we include three different false-alarm probability (FAP) levels and the location of the transiting planet at 2.527\,d as a reference. We also indicate the location of highest peak of the CARMENES RV GLS periodogram for completeness (see Section~\ref{subsec:CARMENES RV data}). In the timeline, the ASAS-SN $V$-band photometry is the oldest dataset, then ASAS-SN $g$-band photometry was taken with about 170\,d overlap with the $V$-band photometry. The TJO $R$-band photometric time series followed just after the ASAS-SN $g$-band data. Finally, the OSN photometry consists of the most recent data, and they overlap for a period of $\text{about one}$ month with the TJO observations. For all light curves, we explored periodicities in the interval 1.1 through 1000 days. 

One of the highest peaks of the ASAS-SN $V$ and $g$-band GLS periodogram (top panels of Fig.~\ref{fig:GLS_phot_multiplot}) occurs at $\sim$27\,d (green band), but it is not very significant. It agrees well with the peak found in the \textit{TESS} light curve analysis, and the ASAS-SN $g$-band photometry is contemporaneous with the \textit{TESS} data. However, this peak is not seen in ASAS-SN $g$-band data taken with the bs camera (third panel of Fig.~\ref{fig:GLS_phot_multiplot}). The GLS periodogram of the TJO $R$-band photometry is rather puzzling (fourth panel of Fig.~\ref{fig:GLS_phot_multiplot}): When the highest peak at less than 2\,d is ignored (the periodicity is too short for the stellar rotation of an M1.5V star), another significant peak is located at $\sim$12\,d. This value is similar to the $P_{\rm 1/2\,rot}$ value obtained with the \textit{TESS} analysis. However, in the TJO GLS periodogram, the region around $\sim$27\,d ($P_{\rm \,rot}$ candidate) is free of significant peaks. In order to determine whether this strong peak in the TJO periodogram at $\sim$12\,d veils other peaks at longer periodicities (i.e., agrees better with the \textit{TESS} and ASAS-SN data), we subtracted the trend observed in the original data and performed a new GLS analysis on the residuals. The resulting GLS periodogram is shown in the fifth panel of Fig.~\ref{fig:GLS_phot_multiplot}: Two peaks gain power and become significant close to the green region (25--30\,d), but they do not correspond to the highest peaks of the periodogram. Given the ratio 2:1 between the two peaks ($\sim$12 and $\sim$27\,d), it is likely that one of the peaks is the first harmonic of the other. The GLS periodograms of the OSN $V$- and $R$-band light curves (two bottom panels of Fig.~\ref{fig:GLS_phot_multiplot}) confirm the $P_{\rm \,rot}$ value of TOI-1470 as they are dominated by a highly significant peak at $\sim$29\,d.

Our conclusion is that from all ASAS-SN, TJO, and OSN photometric data sets covering a total of 5.6\,yr of regular monitoring, it is possible to extract one single characteristic stellar rotation period for TOI-1470 at the value of $P_{\rm \,rot}$=29$\pm$3\,d, where the error bar comes from the width at half height of the highest peak in the GLS. This $P_{\rm \,rot}$ value agrees with that found from \textit{TESS} light-curve analysis in the previous section and is considered the rotation period of the star (see Table~\ref{tab:stellar_properties_TOI-1470}). The stellar variability produced by spots located at different longitudes of the stellar surface or the variations in the spot coverage or sizes in this particular case also revealed the $P_{\rm 1/2\,rot}$ value. For this reason, depending on the configuration of the spots, we detected $P_{\rm \,rot}$ (ASAS-SN case), $P_{\rm 1/2\,rot}$ (TJO case), or both (OSN case).

\subsubsection{CARMENES activity indicators \label{spec-indicators}}

We computed the GLS periodograms of some stellar activity indicators included in the CARMENES {\tt serval} pipeline. The spectroscopic GLS periodograms are shown in panels f--i of Fig.~\ref{fig:GLS_activity_multiplot} in the range of 1.1--350\,d, which is the time coverage of the observations. The top panel of this figure displays the window function of the CARMENES observations, where the most significant peaks are located at 1\,d and close to half a year.

There are three Ca\,{\sc ii} IRT indices, one per atomic line; their corresponding GLS periodograms are depicted together in panel f of Fig.~\ref{fig:GLS_activity_multiplot}. The H${\alpha}$ index periodogram is shown in panel g of the same figure. The periodograms of the Ca\,{\sc ii} IRT and H${\alpha}$ show peaks at $\sim$25\,d that reach FAP levels above 0.1\,\%~(Ca\,{\sc ii} IRT) and 1\,\%~(H${\alpha}$ ) significance levels. In both cases, the peak is quite broad, with a pedestal extending from $\sim$20 through $\sim$40\,d (this is marked with a green band in all panels of Figs.~\ref{fig:GLS_phot_multiplot} and~\ref{fig:GLS_activity_multiplot}).

The GLS periodogram of the CRX and dLW indices (bottom panels of Fig.~\ref{fig:GLS_activity_multiplot}) show no significant peak above any of the three defined FAP levels.  We also investigated correlations between the measured CARMENES RVs and the individual activity indices provided by the {\tt serval} pipeline using Pearson's $r$ and $p$ coefficients to detect correlations and to access the significance of the correlation. We found no strong correlations between the RVs and the activity indices.

In conclusion, the CARMENES Ca\,{\sc ii} IRT and H${\alpha}$ indices suggest that the characteristic frequency for the stellar variations is $\sim$30\,d, which agrees with the results obtained from the analysis of the \textit{TESS} and OSN light curves (Section~\ref{asas-tjo-sno}). Interestingly, there is a prominent peak at this periodicity in the GLS periodogram of the RVs illustrated in the second panel of Fig.~\ref{fig:toi1470_activity_time}, which suggests that the stellar activity has an impact and that this is also seen in the RV data. Because this periodicity ($\sim$30\,d) coincides with that observed from the photometric analysis (from space- and ground-based observations), we ascribe it to the rotation period of TOI-1470 that it is reported in Table \ref{tab:stellar_properties_TOI-1470}.

\section{TOI-1470 planetary system}
\label{sec:analysis_pl}

\subsection{Light-curve transit analysis}
\label{subsec:TESS_light_curve}

\begin{figure}
        \centering
        \includegraphics[width=\columnwidth]{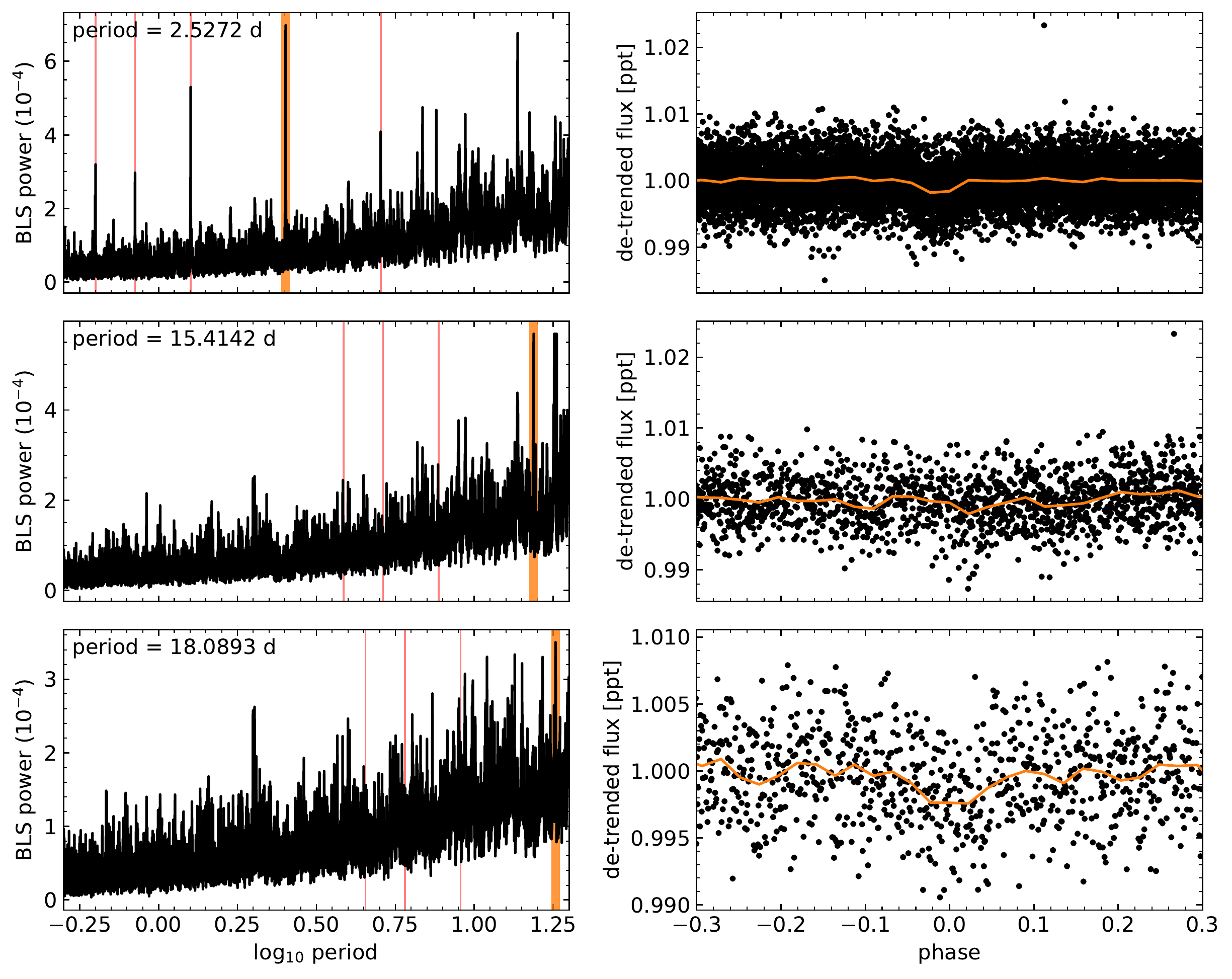}
 
        \caption{BLS periodograms of {\sl TESS} sectors 17, 18, and 24 (left panels) and the photometric data folded in phase (right panels), together with a roughly binned (orange line). Various harmonics are also identified with vertical thin red lines. {\sl Top panel:} Detection of the 2.52d transiting planet (the strongest peak is marked with a vertical orange line). {\sl Middle panel:} Detection of a potential 15.41\,d signal after masking the previous signal. The LC folded in phase does not show any evidence of a transit. {\sl Bottom panel:} Detection of the 18.08d transiting-planet candidate (strongest peak) after masking the two previous signals. The LC data folded in phase appear to indicate a possible shallow transit.}
        \label{fig:BLS_3LCs}
\end{figure}

\textit{TESS LCs}. We applied the box least-squares periodogram \citep[BLS,][]{2002A&A...391..369K, 2016A&C....17....1H} to the first three original PDCSAP \textit{TESS} time-series data to verify the 2.5d planet announced by the \textit{TESS} team and to search for additional transits that may have been missed by the \textit{TESS} pipeline. We employed the BLS algorithm programmed in the {\tt astropy.timeseries python} package and made the process iterative: When a transiting planet candidate was identified, it was masked, and the algorithm was run again to search for additional transiting planets, from the strongest to the weakest signals. The BLS periodograms of the PDCSAP fluxes for the three sectors were computed for periodicities in the range of 0.5--20\,d and are depicted in Fig.~\ref{fig:BLS_3LCs}. The first identified signal in the BLS is 2.52\,d, which corresponds to the transiting planet candidate announced by \textit{TESS}. The second peak appears at 15.4\,d, but it does not exhibit any transit feature when the photometric data are folded in phase. However, when the two previous signals (2.52 and 15.4\,d) are masked and a new BLS periodogram is computed, an interesting signal appears at 18.08\,d in the third BLS periodogram of Fig.~\ref{fig:BLS_3LCs}. This signal shows a promising LC with a shallow feature when folded in phase. However, most important here is that the 18.08\,d signal present in the BLS periodograms of the LCs can also be identified in the CARMENES RV GLS periodogram (second panel of Fig.~\ref{fig:GLS_activity_multiplot}) and corresponds to the half-value of the second transiting-planet candidate announced by \textit{TESS}. No other significant BLS peaks are identified after these three signals {are} properly masked in the LCs. The {\tt python} package that computes the BLS periodograms also determines the first estimates of the time of periastron passage ($t_0$) and the transit depth ($t_{\rm depth}$). These values agree within the error bars with the reported values by the \textit{TESS} team in the case of TOI-1470\,b.

An inspection by eye of the upper and middle panels of Fig.~\ref{fig:SAP_and_PDCSAP} shows that some of the photometric data available in the original SAP \textit{TESS} light curves were removed when the PDCSAP LCs were computed, particularly in sector 24 (BJD= 2,458,956--2,458,958 and 2,458,970--2,458,972). The different number of data points between the SAP and PDCSAP LCs in sector 24 come from the \textit{TESS} pipeline, which masked these dates when corrections of instrumental variations and crowding were applied. The time coverage of each sector of \textit{TESS} is $\sim$27\,d; we therefore expect, if we are fortunate, two transit events in each sector for our planet candidate at 18.08\,d. With the ephemerides of the planet candidates, we can predict the epoch of each transit. In case the of the 18.08\,d signal, we have transits in sectors 17 and 18, but the transit in sector 24 is missed in the PDCSAP fluxes because it falls in a gap of the LC (see Fig.~\ref{fig:SAP_and_PDCSAP}). It is not only important to rescue the lacking data in sector 24 in order to add information from another transit of TOI-1470\,c to our analysis, but by filling in this gap in sector 24, we will also catch the moment at which the two planetary transits overlap. Working with the entire number of available photometric data, especially in sector 24, is crucial for a better characterization of TOI-1470\,c. Therefore, we decided to use all the available data and detrend the SAP fluxes for the three \textit{TESS} sectors with our own procedure. The SAP fluxes in sector 18 at the beginning of the observations show a clear instrumental effect that has to be corrected for very carefully or even has to be masked. Moreover, the planetary transits were masked before the LC detrending. Our detrended SAP fluxes that we used in our combined analysis for the planetary characterization (Sect. \ref{sec:planet orbiting toi-1470}) are shown in the bottom panel of Fig. \ref{fig:SAP_and_PDCSAP}.

MuSCAT \textit{LCs}. The transit observations with MuSCAT (2.5\,d signal), MuSCAT2, and MuSCAT3 (18.08\,d) in the different filters are affected by instrumental and airmass effects. We corrected the LCs for these effects before fitting. The transits of MuSCAT and MuSCAT2 in the $g'$ band are not good enough and have a large dispersion also after the correction, and we did not take them into account in the final fit. The individual transit analyses of \textit{TESS}, MuSCAT, MuSCAT2, and MuSCAT3 for the two transit signals (2.5 and 18.08\,d) give consistent transit parameters within the error bars, such as the transit depth, the duration, and the ephemerides.

\begin{figure*}[]
        \centering
        \begin{minipage}{0.49\linewidth}
                \includegraphics[angle=0,scale=0.19]{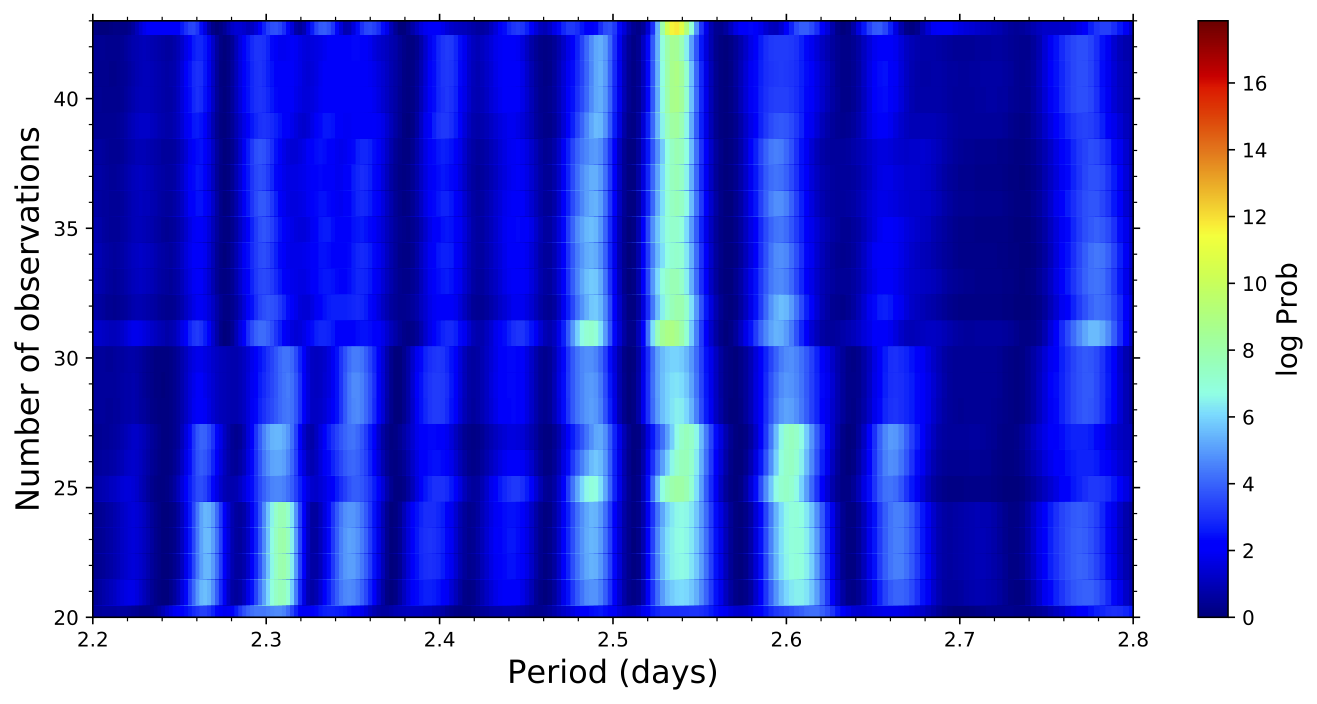}
        \end{minipage}
        \begin{minipage}{0.49\linewidth}
                \includegraphics[angle=0,scale=0.19]{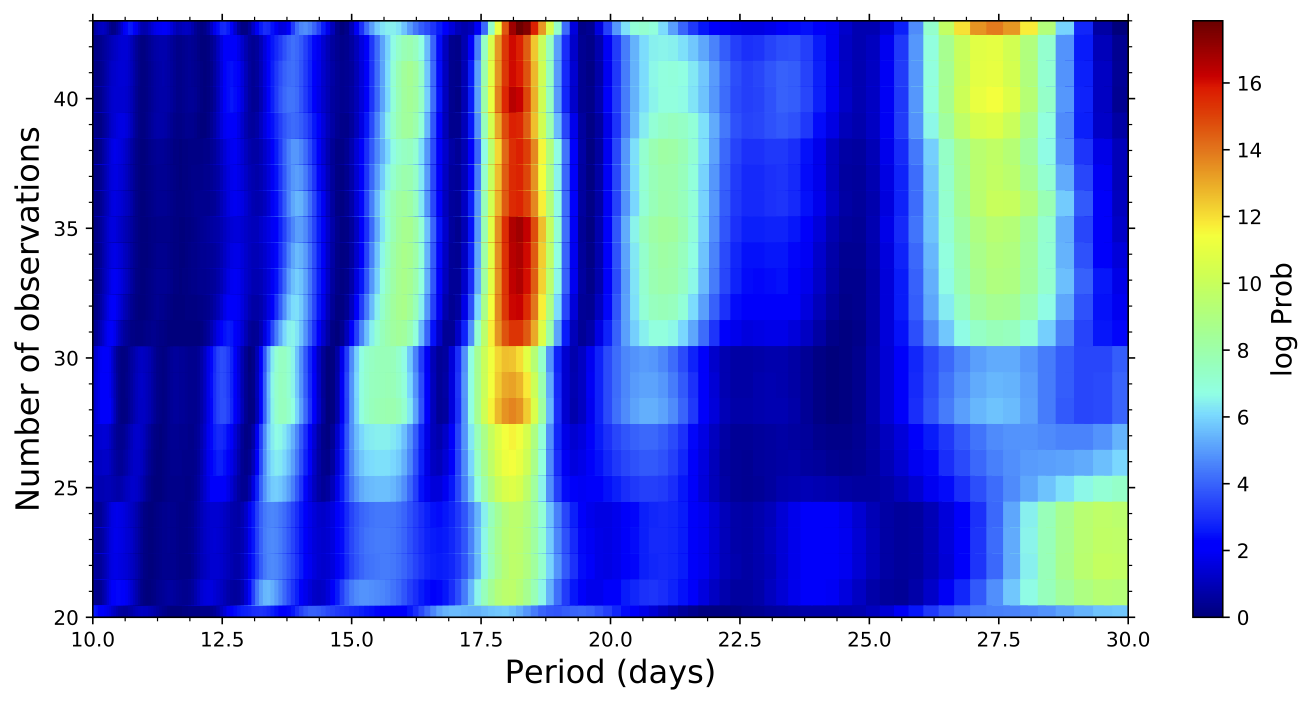}
        \end{minipage}
        \caption{Evolution of the s-BGLS periodogram of the CARMENES RV data around 2.5\,d ({\em left}) and in the region between 10--30\,d ({\em right}).}
        \label{fig:toi1470_color_s-BGLS}
\end{figure*}

\subsection{Limb-darkening coefficients}
\label{subsec:limb-dark}

The limb-darkening coefficients can be determined with {\tt juliet} \citep{2019MNRAS.490.2262E} through the parameterization of $q_1$ and $q_2$ by letting the coefficients vary freely between zero to one \citep{2013MNRAS.435.2152K}. However, we computed the stellar limb-darkening effect through the $u_1$ and $u_2$ coefficients using \texttt{ExoTETHyS}\footnote{\url{https://github.com/ucl-exoplanets/ExoTETHyS}} \citep{morello2020,morello2020joss} with the stellar parameters from Table \ref{tab:stellar_properties_TOI-1470}, and based on the \texttt{PHOENIX} grid of stellar spectra \citep{2009A&A...506.1367W,2013A&A...553A...6H} and the new method for stellar limb darkening computations published by \cite{claret2018}. With the obtained values of $u_1$ and $u_2$ , we transformed them into $q_1$ and $q_2$ and adopted Gaussian priors centered at the obtained values together with their error bars in our final model.


\subsection{CARMENES radial velocity analysis}
\label{subsec:CARMENES RV data}

\subsubsection{GLS periodograms}
\label{subsubsec:CARMENES pre-whitening}

To identify possible aliasing phenomena in the CARMENES RV data caused by the gaps in the time sampling of the observations \citep[e.g.,][]{2010ApJ...722..937D,2020A&A...636A.119S}, we took the spectral window displayed in the top panel of Fig.~\ref{fig:GLS_activity_multiplot} into account. The strong peaks of the window function may introduce alias peaks in the RV periodogram at frequencies according to the expression $f_{\rm alias}$ = $f_{\rm true} \pm m f_{\rm window}$, where $m$ is an integer, $f_{\rm true}$ is the frequency identified in the RV periodogram, and $f_{\rm window}$ is the frequency from the window function \citep{1975Ap&SS..36..137D}. Typical aliases affecting ground-based observations are associated with the year, synodic month, sidereal day, and solar day. In our spectroscopic window function, the highest peaks occur at $\sim$1 d, $\sim$19 d, close to half a year, and beyond 200\,d. We checked that these false signals do not introduce a strong peak in the RV periodogram that can be misinterpreted as being of Keplerian nature. The only exception is a peak at around 70\,d in the CARMENES RV periodogram (panel b of Fig.~\ref{fig:GLS_activity_multiplot}), which might be an alias of the stellar rotation period caused by the $\sim$19d peak of the window function. We remark that the significance of the 70d signal disappears from the CARMENES RV periodogram when the activity signal of the star with a characteristic period of $\sim$28\,d is removed from the data (panel e of Fig.~\ref{fig:GLS_activity_multiplot}).

In the CARMENES RV periodogram, we first subtracted the peak at 2.52\,d (panel c of Fig.~\ref{fig:GLS_activity_multiplot}), and the signals at 18.08\,d and $\sim$28\,d were still there. Then we subtracted the peak at 18.0\,d (panel d of Fig.~\ref{fig:GLS_activity_multiplot}), and the signal associated with the stellar rotation period was still present in the residuals. Finally, we subtracted the signal at $\sim$28\,d (panel e of Fig.~\ref{fig:GLS_activity_multiplot}) that is related to stellar activity signals, and the GLS presented a power at the level of the noise without significant signals. The highest signals observed at $\sim$70\,d in the original CARMENES RV GLS (panel b of the figure) also disappeared when we subtracted the stellar rotation period. From now on, having identified the 2.52\,d and 18.08\,d signal in both photometric and spectroscopic analysis separately, we can clearly note the 2.52\,d and the 18.08\,d signals as confirmed transiting planets: TOI-1470\,b and TOI-1470\,c, respectively. The final orbital parameters of the two planets are determined in Section \ref{sec:planet orbiting toi-1470} with the combined {\sl TESS}--CARMENES data analysis.

We confirm that 18.08\,d is the true orbital period of planet TOI-1470\,c as opposed to the 36.18\,d announced by {\sl TESS}, not only because the CARMENES RVs show a significant peak at the shorter period in the periodogram, but also because all MuSCAT2 and MuSCAT3 observations were designed with the 18.08d ephemeris, and we were able to recover the planetary transits in all attempts. This also ensures that the derived ephemeris parameters are reliable.

We verified that the RV signals at 2.52\,d and 18.08\,d are stable and coherent over the entire observational time baseline by producing the stacked Bayesian generalized Lomb-Scargle periodogram \citep[s-BGLS,][]{2015A&A...573A.101M} shown in Fig.~\ref{fig:toi1470_color_s-BGLS}. The significance or probability of both planetary signals increases with time until a stable level is reached at a certain number of observations. This is more evident for the 18.08\,d signal. Then, the signals become narrower. This behavior is expected for signals with a Keplerian origin. This further supports the exoplanetary nature of TOI-1470\,b and c. In contrast, the signal between 26--32\,d, where we placed the value of stellar rotation period ($P_{\rm rot} = 29\pm3$\,d), is more erratic(as expected for signals with a stellar origin). It changes its period value when the number of observations is increased.

\subsubsection{Radial velocity models}
\label{sec:radvelmodel}

\begin{table}[]
\centering
        \begin{small}
                \centering
                \caption{Comparison of different {\tt juliet} RV models for TOI-1470 using the CARMENES RV data.}
                \label{tab:toi1470_rv_model_comparison}
                \begin{tabular}{l l l c c}
                        
                        \hline
                        \hline
                        \noalign{\smallskip}
                        Model  & Description & $\ln \mathcal{Z}$ \\
                        \noalign{\smallskip}    
                        \hline  
                        \noalign{\smallskip}
                        BM                          & RV offset and jitter                  & --157.3 \\ 
                        
                        1pl                         &                                  & --152.7 \\ 
                        
                        2pl                         &                                   & --149.3 \\ 
                        
                        
                        2pl-ecc                         &                                       & --155.4 \\
                        
                        1pl$+$GP                            &          $P_{\rm rot,GP} \sim$ 20--50\,d                                 & --151.3 \\
                        
                        
                        2pl$+$GP                            &   $P_{\rm rot,GP} \sim$ 20--50\,d          & --150.4 \\ 

                                
                        \noalign{\smallskip}    
                        \hline  
                        \noalign{\smallskip}
                \end{tabular}
        \end{small}
\end{table}

We modeled the CARMENES RV time series in order to constrain the most critical orbital parameters for a subsequent more sophisticated joint photometric and spectroscopic analysis of the TOI-1470 planetary system (presented in Section~\ref{sec:planet orbiting toi-1470}). This step was necessary to save computing time. The models were produced with the {\tt juliet} code, which calls the {\tt radvel} \citep{2018PASP..130d4504F} package to model Keplerian RV signals. The stellar activity signals were modeled by means of a Gaussian process (GP) with a quasi-periodic kernel by \cite{2017AJ....154..220F} and provided by {\tt celerite} python library, which is suited for learning periodic functions:

\begin{equation}
        \label{eq:QPK_kernel}
        k_{i,j}(\tau) = \frac{B}{2+C} e^{-\tau/L} \left[ \cos \left( \frac{2\pi \tau}{P_{\rm rot}} \right) + (1 + C) \right],
\end{equation}

\noindent where $\tau= \abs{t_{i}-t_{j}}$ is the time lag, $B$ and $C$ are parameters related to the amplitude of the GP, $L$ is related to the timescale for the amplitude-modulation of the GP, and $P_{\rm rot}$ is the period of the quasi-periodic modulations. In order to simplify the GP equation, we fixed the $C$ parameter to 0.

We based the selection of the best model on the rules defined by \cite{2008ConPh..49...71T} for the Bayesian model log evidence, $\ln{\mathcal{Z}}$: if $\Delta \ln{\mathcal{Z}} \le 3,$ the two models are indistinguishable and neither is preferred, while if $\rm \Delta \ln{\mathcal{Z}} > 3,$ the model with the larger Bayesian log evidence is favored. We performed four different approaches that are summarized in Table~\ref{tab:toi1470_rv_model_comparison}. All included one base model (BM) consisting of RV offset and jitter. The other ingredients are the following:
\begin{itemize}
        \item One Keplerian signal at 2.52\,d (1pl model).
        \item Two Keplerian signals at 2.52\,d and 18.08\,d (2pl model). This model was explored with both a zero eccentricity and a free eccentricity for planets TOI-1470\,b and c (2pl-ecc model). The computations indicate that an eccentricity near zero is preferred.
        \item One Keplerian signal at 2.52\,d plus a periodic GP with $P_{\rm rot}$ in the interval 20--50\,d to simulate the stellar activity due to stellar rotation (1pl$+$GP).
        \item Two Keplerian signals at 2.52\,d (TOI-1470\,b) and 18.08\,d (TOI-1470\,c) plus a periodic GP with $P_{\rm rot}$ in the interval 20--50\,d to simulate the stellar activity due to stellar rotation (2pl$+$GP).
\end{itemize}

The resulting log evidence for each model is provided in Table~\ref{tab:toi1470_rv_model_comparison}. When only the RV information is used, four models are equally probable: 1pl, 2pl, 1pl$+$GP, and 2pl$+$GP. The 2pl$+$GP model is physically the most plausible because it includes all observed components that were previously explained in the data (two transiting planets and the stellar rotation).

\subsection{Masses of the transiting planets}
\label{sec:planet orbiting toi-1470}

\begin{table}[]
        \renewcommand{\arraystretch}{1.3}
                \centering
                \caption{Final adopted planetary parameters for the TOI-1470 system.}
                \label{tab:toi1470bc_params_from_joit-fit}
                \begin{tabular}{l c c }
                        
                        \hline
                        \hline
                        \noalign{\smallskip}
                        Parameter         &             TOI-1470\,b & TOI-1470\,c  \\
                        \noalign{\smallskip}
                        \hline
                        \noalign{\smallskip}
                        
                        \noalign{\smallskip}
                        \multicolumn{3}{c}{\textit{Fitted planet parameters}} \\
                        \noalign{\smallskip}
                        $P$  (d)                                                                &$2.527093^{+0.000004}_{-0.000003}$ &  $18.08816^{+0.00006}_{-0.00008}$ \\
                        $t_0$ $^{(1)}$                          &$1766.4702^{+0.0006}_{-0.0006} $ & $1772.176^{+0.003}_{-0.002}$ \\
                        $e$                                                                     &  $\le 0.3 $ & $\le 0.5$  \\
                        $K$ (m/s)                                                       &   $5.67^{+0.92}_{-0.96}$ & $2.91^{+1.15}_{-1.11}$  \\
                        $r_{1}$                                                         &   $0.72^{+0.02}_{-0.02}$ & $0.63^{+0.02}_{-0.01}$  \\
                        $r_{2}$                                                         &   $0.0426^{+0.0008}_{-0.0007}$ & $0.0481^{+0.0004}_{-0.0004}$  \\

                        \noalign{\smallskip}
                        \multicolumn{3}{c}{\textit{Derived planet parameters} } \\
                        \noalign{\smallskip}
                        $R_{p}/R_{\star}$                               &  $0.0426^{+0.0008}_{-0.0007}$ & $0.0481^{+0.0004}_{-0.0004}$  \\
                        $R_{\rm p}$ (R$_{ \oplus}$)             & $2.18^{+0.04}_{-0.04}$ & $2.47^{+0.02}_{-0.02}$   \\
                        $a/R_{\star}$                                                   &  $13.05^{+0.14}_{-0.15}$ & $48.46^{+0.52}_{-0.54}$   \\
                        $a$ (au)                                                                                                        &$0.0285^{+0.0004}_{-0.0004}$ & $0.106^{+0.001}_{-0.001}$ \\
                        $b = (a/R_{\star}) \cos i$              &  $0.59^{+0.02}_{-0.02}$ & $0.47^{+0.02}_{-0.02}$   \\
                        $i$ (deg)                                                                       &  $87.42^{+0.12}_{-0.12}$ & $89.47^{+0.03}_{-0.03}$   \\
                        
                        $t_{\rm 14}$ (h)                                                & $1.28^{+0.02}_{-0.02}$ & $2.7^{+0.01}_{-0.01}$   \\
                        $t_{\rm depth}$ (ppm)                   & $1814^{+65}_{-61}$ & $2315^{+34}_{-34}$  \\
                        
                        $M_{\rm p}$ (M$_{ \oplus}$)                                     & $7.32^{+1.21}_{-1.24}$ & $7.24^{+2.87}_{-2.77}$   \\
                        $\rho_{\rm p}$ (g $\rm cm^{-3}$)                & $3.86^{+0.70}_{-0.68}$ & $2.66^{+1.06}_{-1.02}$  \\
                        $T_{\rm eq}$ (K)$^{(2)}$                                &  551--734 &  287--373  \\
                        $S$ (S$_{\oplus}$)                 & $46.3^{+1.5}_{-1.4}$ & $3.35^{+0.3}_{-0.2}$ \\
                        
                        \noalign{\smallskip}
                        \hline
                \end{tabular}
                \tablefoot{$^{(1)}$ $t_0$ (BJD $-$ 2,457,000). $^{(2)}$ For the Bond albedo in the interval 0.65--0.0. }
\end{table}

\begin{figure*}[]
        \centering
        \includegraphics[width=0.47\textwidth]{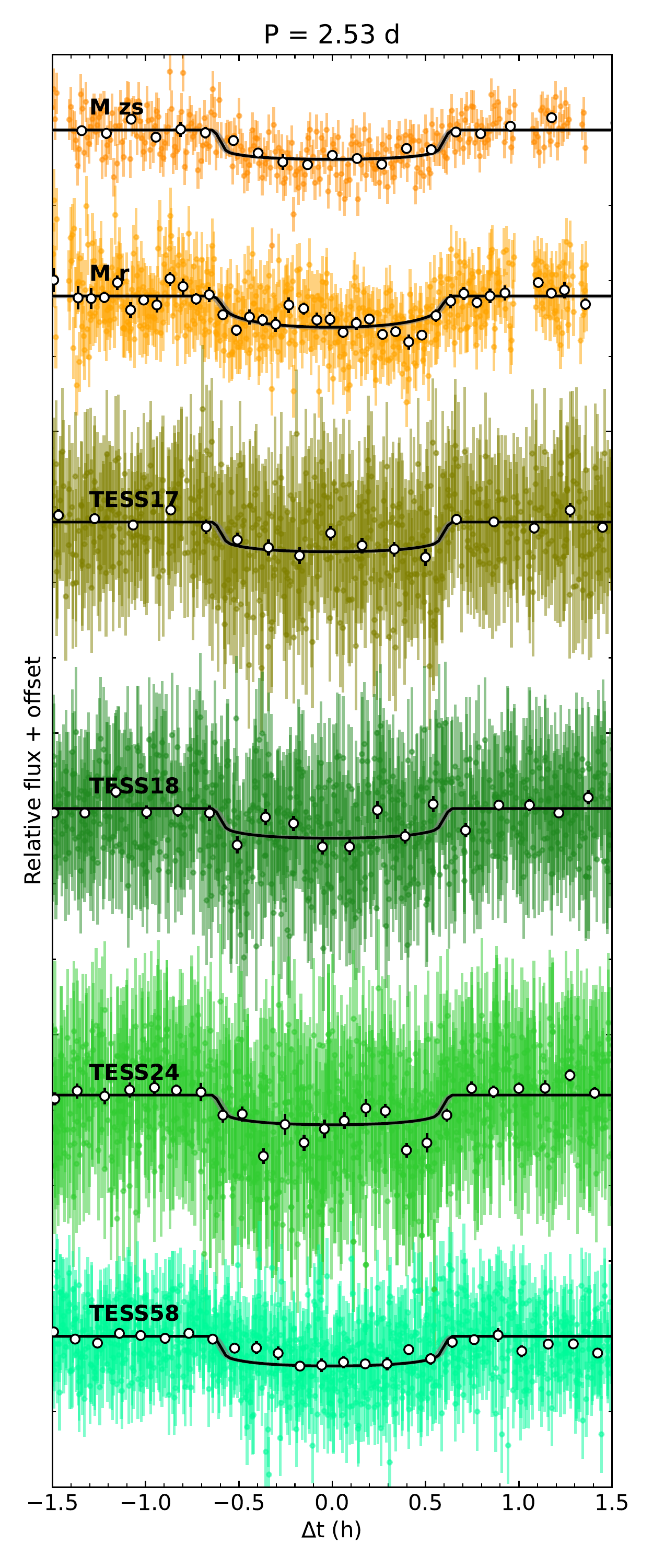}
        \includegraphics[width=0.47\textwidth]{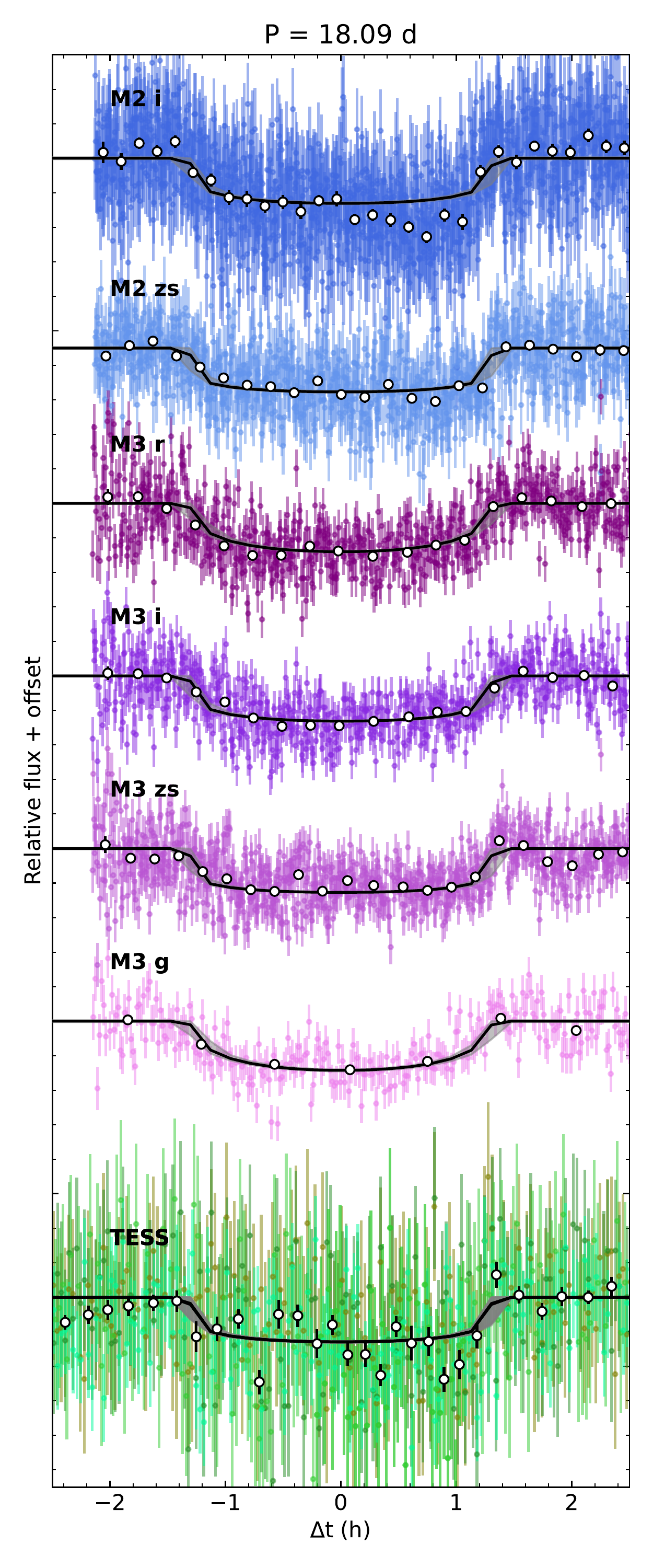}
        \caption{Individual light curves folded in phase with the orbital periods of the transiting planets per filter or sector (colored points). The best joint fit is plotted as a black line. The white dots correspond to the binned photometric data. The x-axis represents the time computed from the mid-transit times as derived from the best joint fit.}
        \label{fig:lc_vs_phase_vertical}
\end{figure*}

\begin{figure}[]
        \centering
        \includegraphics[width=\columnwidth]{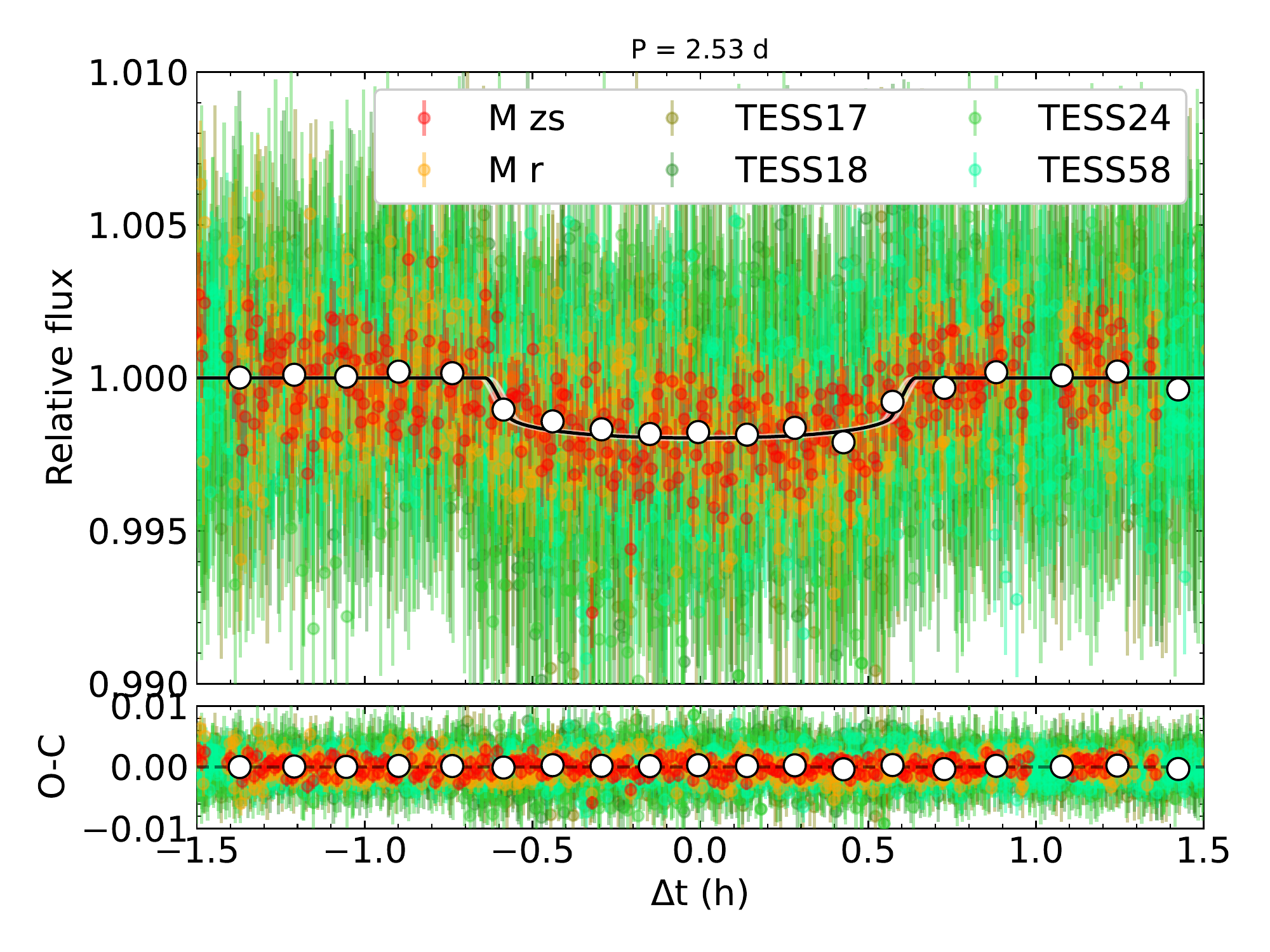}
        \includegraphics[width=\columnwidth]{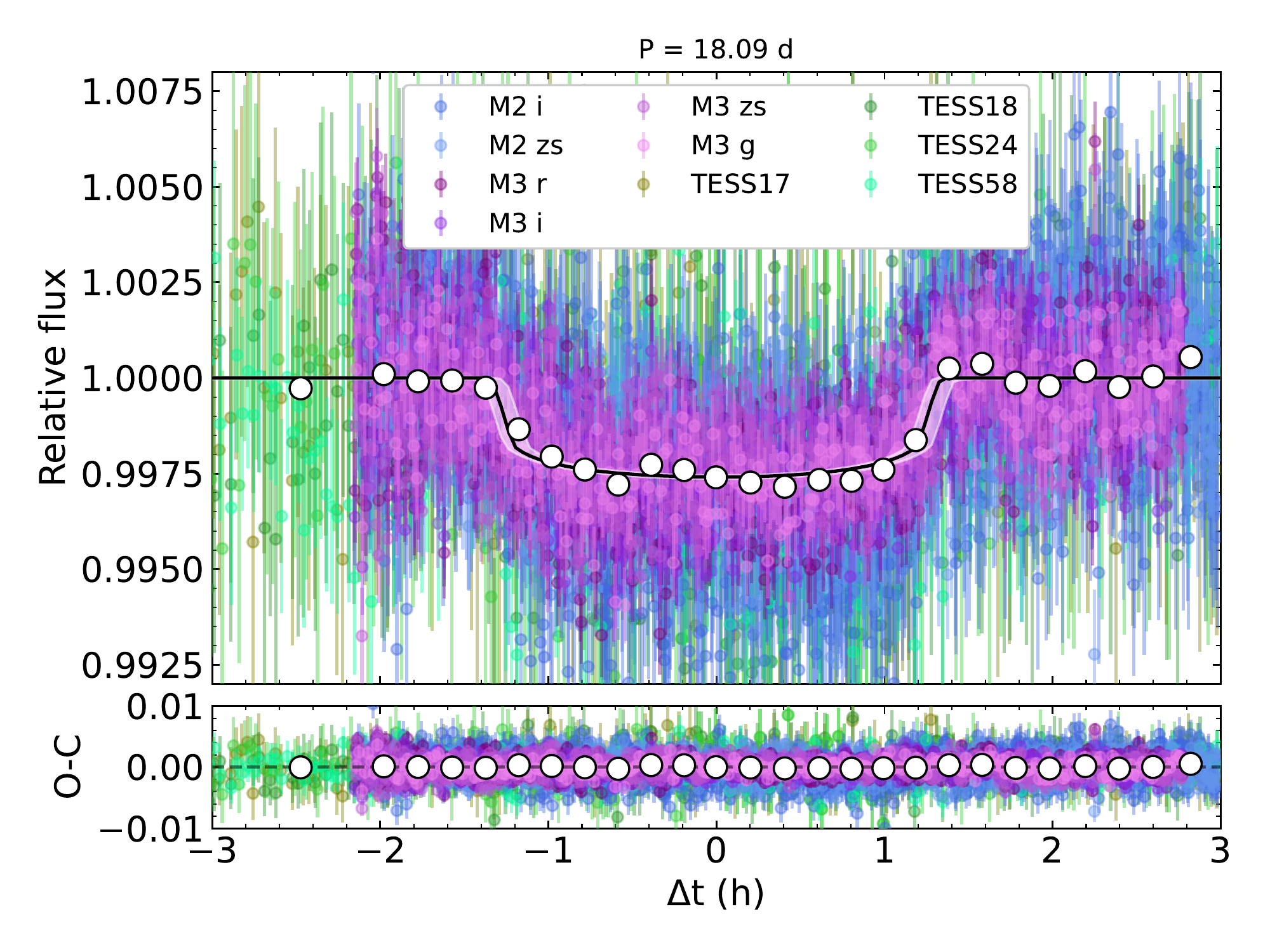}
        \caption{All light curves and the residuals folded in phase with the orbital periods of the transiting planets. The x-axis represents the time from the mid-transit times as derived from the best joint fit. The black line corresponds to the \textit{TESS} transit model and shows the pattern of the limb darkening for \textit{TESS} bandpass. The white dots correspond to the binned photometric data. }
        \label{fig:lc_vs_phase_ALL}
\end{figure}

\begin{figure}[]
        \centering
        \includegraphics[width=\columnwidth]{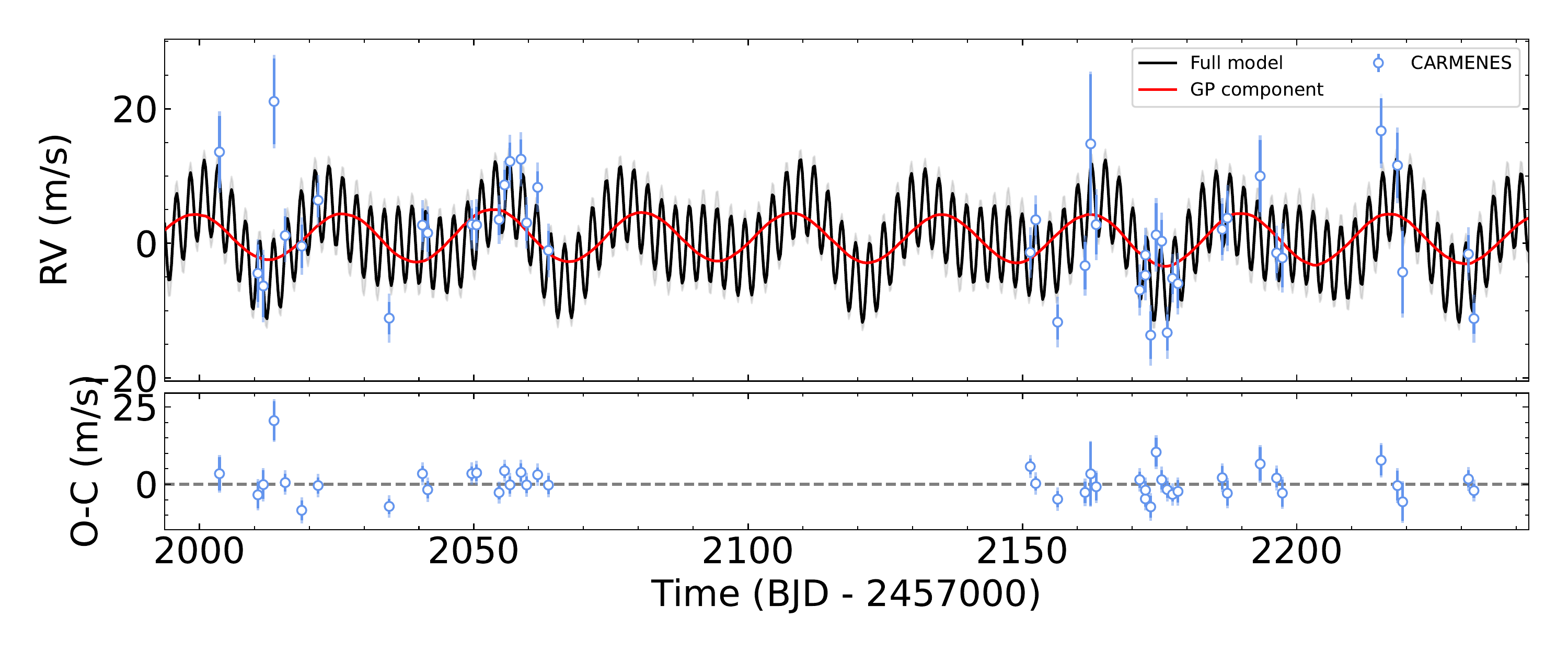}
        \caption{TOI-1470 CARMENES RVs (blue dots) and the best model (black line) with its 1$\sigma$ confidence level (gray shaded area) obtained from the combined photometric and spectroscopic fit. The top panel shows the entire RV time series as a function of the time. The red line shows the GP component that models the stellar activity. The bottom panel shows the RV residuals after subtracting the full model. All error bars include the quoted CARMENES uncertainties and the RV jitter as obtained from the model added in quadrature.
        }
        \label{fig:toi1470_rv_time}
\end{figure}

\begin{figure}[]
        \centering
        \includegraphics[width=\columnwidth]{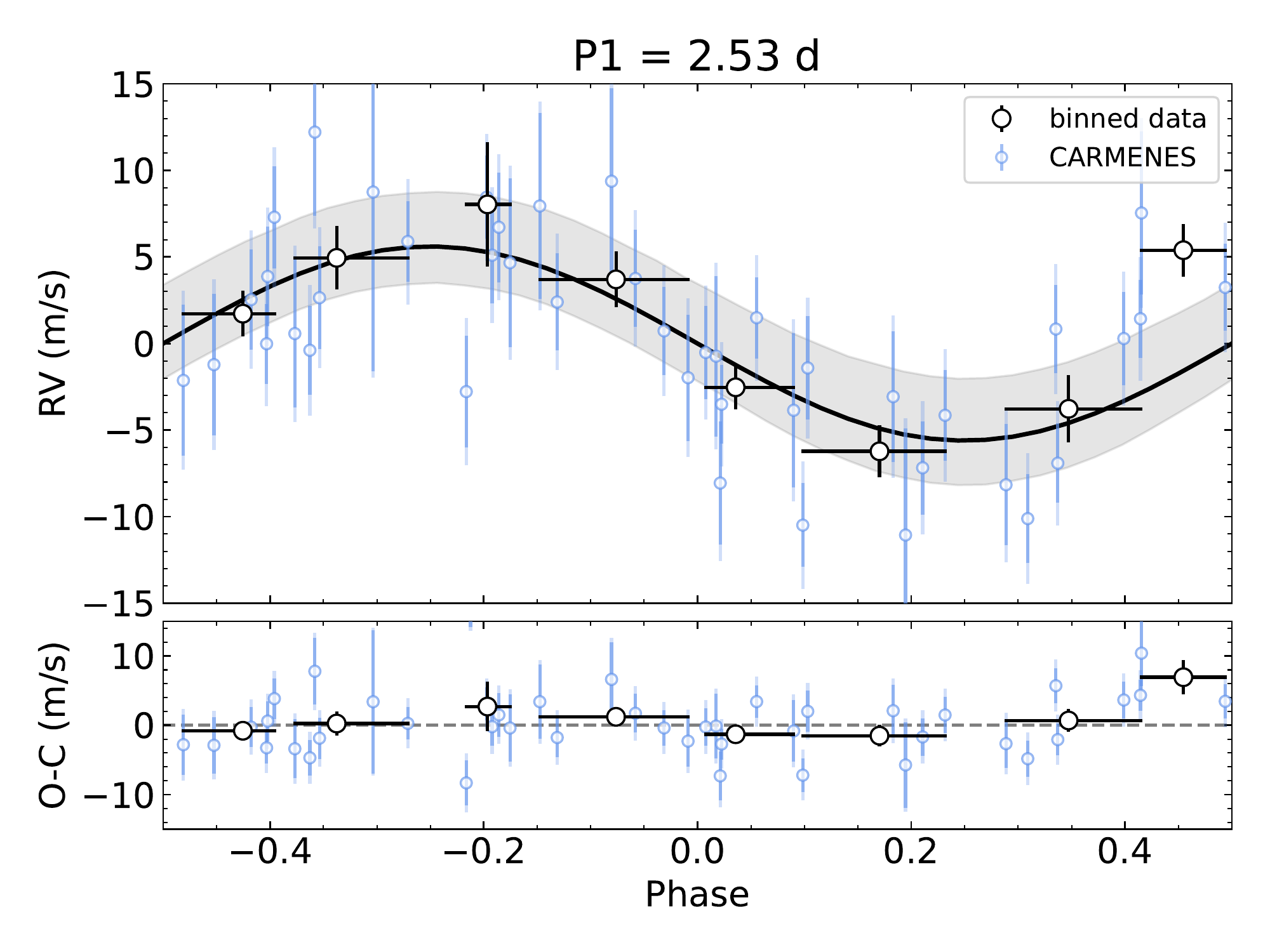}
        \includegraphics[width=\columnwidth]{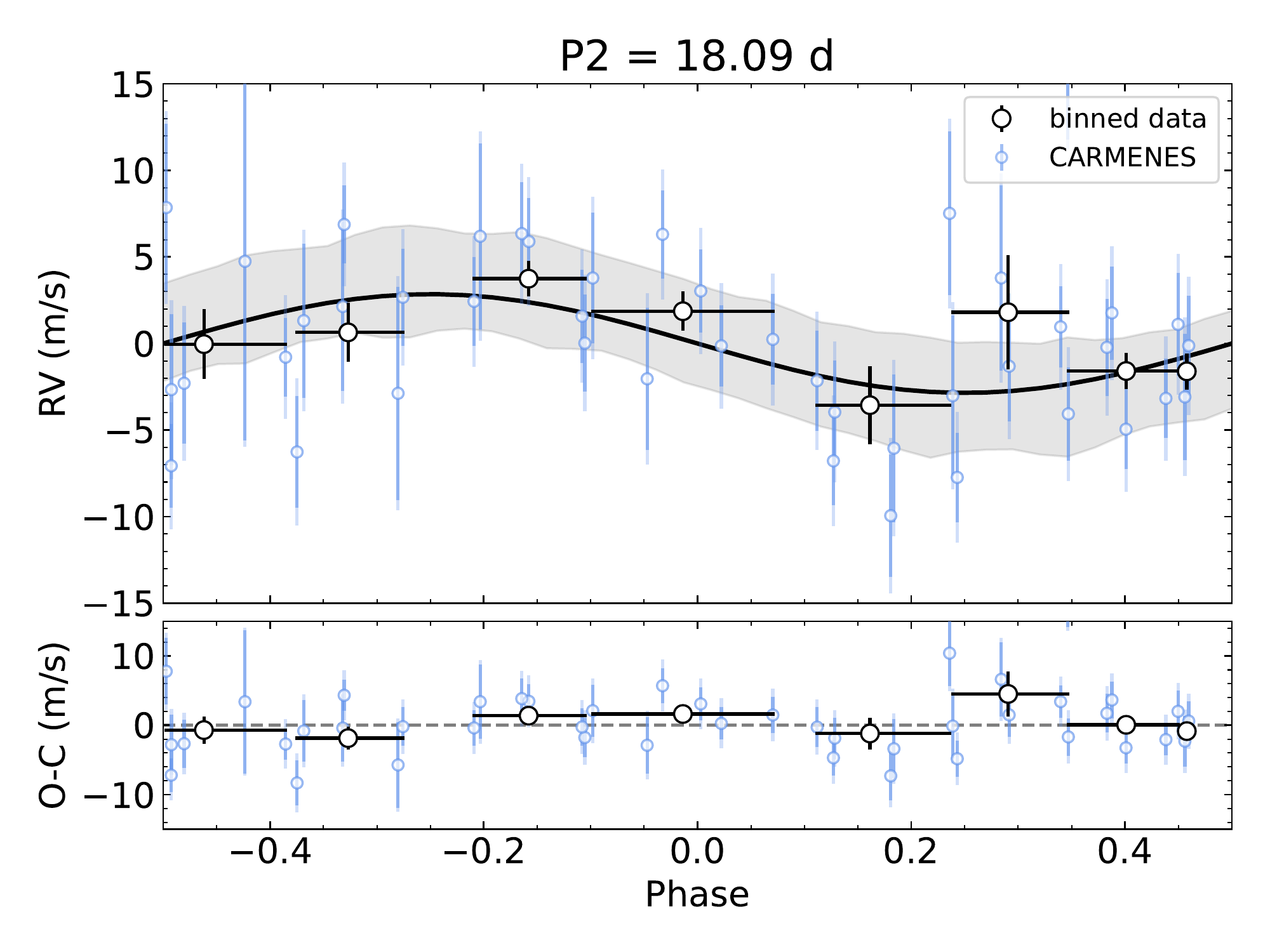}
        \caption{TOI-1470 CARMENES RVs (blue dots) and the best model (black line) from the combined photometric and spectroscopic fit folded in phase with the orbital period of the Keplerian components: TOI-1470\,b and c. The GP component has been removed from the data. The binned data are plotted as open black symbols. All error bars include the quoted CARMENES uncertainties and the RV jitter as obtained from the model added in quadrature. The gray shadowed area denotes the 68\% confidence interval.}
        \label{fig:toi1470_rv_phase_plots}
\end{figure}

To determine the masses of TOI-1470\,b and c, we performed a combined photometric and spectroscopic analysis using the \textit{TESS}, MuSCAT, MuSCAT2, and MuSCAT3 data corrected by us and the CARMENES VIS data. The corrected LCs for each planet and filter folded in phase are presented in Fig.~\ref{fig:lc_vs_phase_vertical}. Our final model consists of two transiting planets at 2.52 and 18.08\,d and the stellar rotation component at 20--50\,d. The first two components were modeled with two circular orbits, whereas the stellar activity component was modeled with the quasi-periodic GP kernel used in the previous section. The GP provides a better model of the stellar activity than a simple sinusoidal function (e.g., \citealt{2021A&A...645A..58P}). We used the medians obtained from the transit-only analysis (Section~\ref{subsec:TESS_light_curve}) to define normal priors on the orbital period and time of periastron passage of the transiting planets. This is fully justified because these parameters are mainly constrained by the light curves and the RV data add little information \citep{2020A&A...642A.236K}. The RV amplitudes ($K$) were fit by adopting a prior with a uniform distribution for each Keplerian signal. We also fit jitters and offsets for the \textit{TESS}, MuSCAT, MuSCAT2, MuSCAT3, and CARMENES data. Finally, we adopted normal distributions to fit the limb-darkening coefficients centered at the values derived by us and explained in detail in Sect. \ref{subsec:limb-dark}. The dilution factor was fixed to one, assuming that no other source in the field introduces a photometric signal that contaminates the LCs. All priors are summarized in Table~\ref{tab:1470_priors_details}.

The posteriors from the fit that we adopted as planetary parameters for the TOI-1470 system are presented in Table~\ref{tab:toi1470bc_params_from_joit-fit}. For clarity, the posteriors of the remaining fit parameters can be found in Table~\ref{tab:toi1470bc_params_from_joit-fit_rest}.

The derived GP period using the RV data is $27.3^{+0.6}_{-0.5}$\,d, which agrees within the quoted uncertainties with the stellar rotation period determined separately from the \textit{TESS} light curves, photometric ground-based data, and from the CARMENES spectroscopic activity indicators. The light curves folded in phase with the orbital periods of the transiting planets with all sectors combined are illustrated in Fig.~\ref{fig:lc_vs_phase_ALL}. For completeness, the corner plot depicting all the posterior distributions of the planetary parameters as obtained from the joint fit is shown in Fig.~\ref{fig:toi1470_cornerplot}.

The resulting RV model is depicted in Fig.~\ref{fig:toi1470_rv_time}, and the RV curves folded in phase for the two transiting planets are shown in Figure \ref{fig:toi1470_rv_phase_plots}. The $rms$ of the RV residuals (i.e., observed RVs minus the best fit) is 5\,m\,s$^{-1}$, which is very similar to but slightly higher than the mean value of the CARMENES VIS RV errors (see Section~\ref{subsec:carmenes_spectroscopy}). This suggests that there is no other component at the level of the noise in our data. With RV amplitudes of 5.67$^{+0.92}_{-0.96}$\,m\,s$^{-1}$ (TOI-1470\,b) and 2.91$^{+1.15}_{-1.11}$\,m\,s$^{-1}$ (TOI-1470\,c), these transiting planets have true masses of 7.32$^{+1.21}_{-1.24}$\,M$_\oplus$ and 7.24$^{+2.87}_{-2.77}$\,M$_\oplus$, respectively, with a significance of 6 and 3\,$\sigma$. 

We also explored whether the two planets have eccentric orbits by leaving this parameter free in our simulations. The results revealed that the eccentricities of TOI-1470\,b and c are poorly constrained and consistent with zero, and we derived an upper limit on the orbital eccentricities of $e \le 0.3$ and 0.5 at the 1\,$\sigma$ level for planets b and c, respectively. 

In summary, TOI\,1470 is the host of two sub-Neptune planets, TOI-1470\,b, with a short orbital period (2.52\,d), and TOI-1470\,c, with a longer orbit (18.08\,d). We also derived the bulk densities of the transiting planets, $\rho = 3.86^{+0.70}_{-0.68}$ and 2.66$^{+1.06}_{-1.02}$\,g\,cm$^{-3}$ for planets b and c, respectively.

\section{Discussion}
\label{sec:discussion}

\begin{figure}[]
        \centering
        \includegraphics[width=\columnwidth]{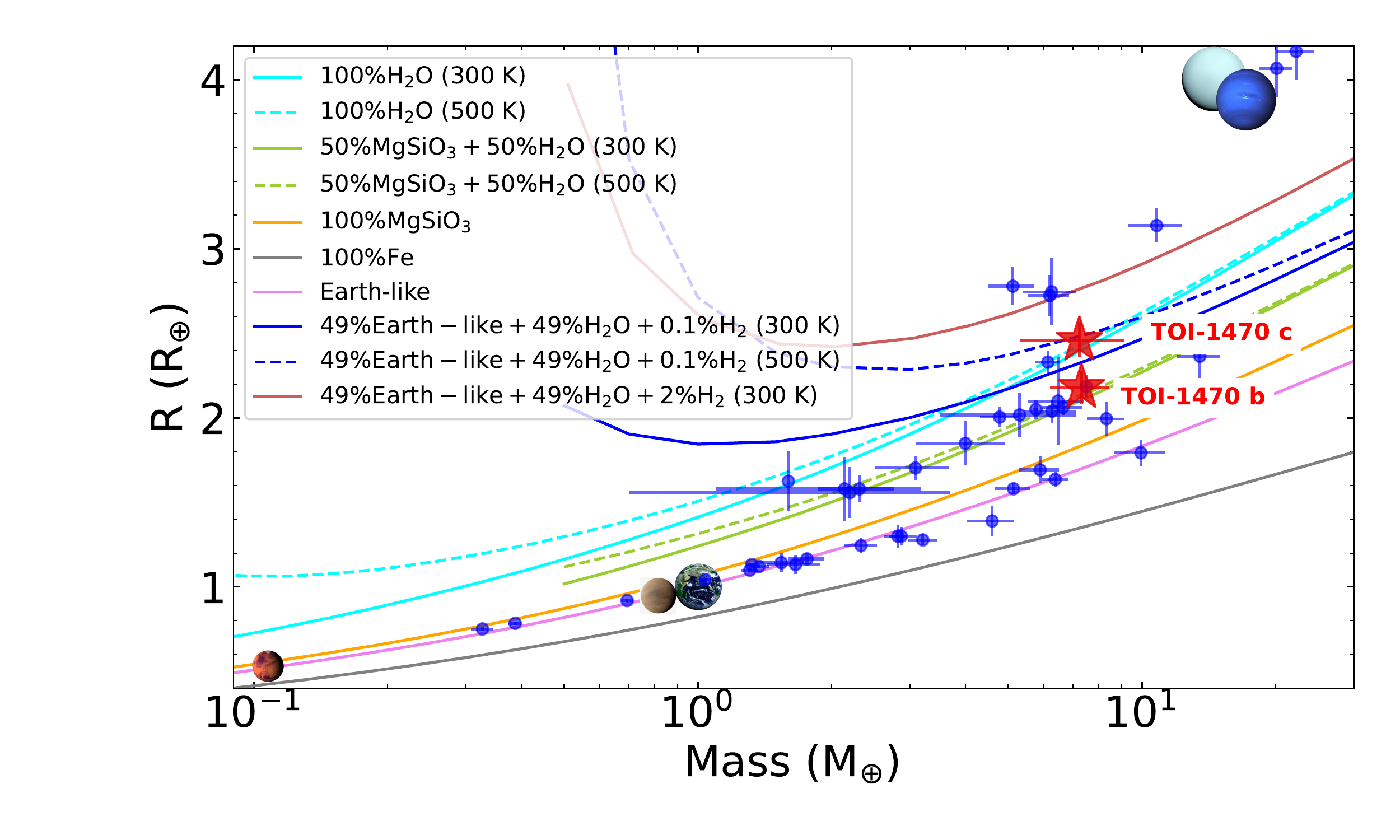}
        \caption{Planetary radius -- mass diagram obtained for planets orbiting M-type stars (blue dots) and discovered via the RV and transit methods (based on the NASA Exoplanet Archive, May 2023). All planets have masses lower than 14\%\ and radii uncertainties smaller than 10\,\%. The various planetary composition models of \cite{2016ApJ...819..127Z} are shown with solid colored lines. The red stars indicate the locations of TOI-1470\,b and c. Venus, Earth, Mars, Jupiter, and Uranus are also shown for comparison purposes.}
        \label{fig:toi1470_planets_around_dM}
\end{figure}

\begin{figure}[]
        \centering
        \includegraphics[width=\columnwidth]{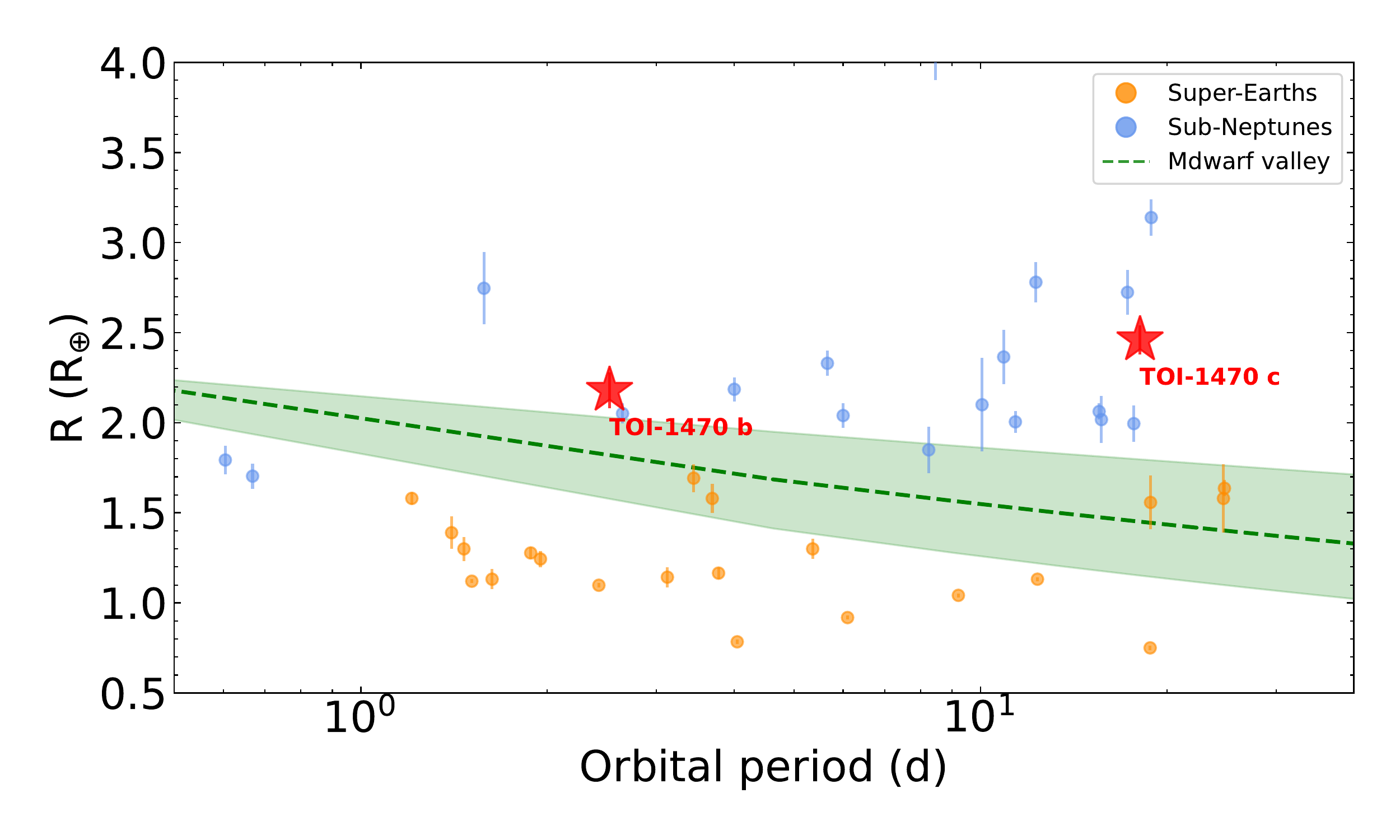}
        \caption{Radius as a function of planetary orbital period (same exoplanet sample as in Figures~ \ref{fig:toi1470_planets_around_dM}). The two transiting planets of the TOI-1470 system are marked with a red star. The dashed green line indicates the radius valley of planetary systems with M-type parent stars.}
        \label{fig:toi1470_planets_radius_valley}
\end{figure}

\cite{2014ApJ...787L..29K} calculated a conservative estimate of the inner habitable zones (HZ) around stars with effective temperatures in the range 2600--7200\,K for planetary masses between 0.1 and 5\,M$_\oplus$. According to these authors, the conservative inner edge of the HZ around TOI-1470 is located at 0.2\,au with an effective stellar flux incidence of $S_{\rm eff} =0.94\,S_{\odot}$. TOI-1470\,b and~c lie between the star and the inner boundary of the HZ. The theoretical equilibrium temperature ($T_{\rm eq}$) of these two  planets was derived by using the Stefan–Boltzmann equation, the stellar parameters given in Table~\ref{tab:stellar_properties_TOI-1470}, and two very different values of the planetary albedo ($A = 0.65$ and~0.0 for the high- and nonreflectance cases). The resulting $T_{\rm eq}$ ranges are 551--734\,K and 287--373\,K for TOI-1470\,b and c, respectively, and are listed in Table~\ref{tab:toi1470bc_params_from_joit-fit}.

To place the TOI-1470 planetary system in context with other known systems with M-type parent stars, we produced Figs.~\ref{fig:toi1470_planets_around_dM} and \ref{fig:toi1470_planets_radius_valley} by using the NASA Exoplanet Archive and planets discovered via the combined RV and transit methods. Both figures show well-characterized planets with masses and radii determined to better than 14\% and 10\%, respectively. Several studies have confirmed the existence of a planetary radius valley at around 1.6\,R$_{\oplus}$ \citep[e.g.,][]{2017AJ....154..109F, 2018MNRAS.479.4786V}, which is illustrated in the planetary radius versus orbital period diagram of Fig.~\ref{fig:toi1470_planets_radius_valley}. This figure depicts the radius valley determination by \cite{2021MNRAS.507.2154V}. On one side of the valley lie small planets (typically below 2\,R$_{\oplus}$) with very tiny or no atmospheres, the so-called super-Earths, while on the other side lie planets with inflated He-H envelopes, the so-called mini(sub)-Neptunes. The two planets of the TOI-1470 system lie on the sub-Neptunes domain, as illustrated in Fig.~\ref{fig:toi1470_planets_radius_valley}. However, the innermost planet TOI-1470\,b is one of the few short-orbital-period sub-Neptunes closest to the radius valley. This may suggest that if TOI-1470\,b were to have a thin atmosphere that is exposed to high irradiation, it might experience photoevaporation processes, implying a planetary mass loss and a size reduction that would move the planet to the regime of stripped rocky cores. Figure~\ref{fig:toi1470_planets_around_dM} shows the planetary mass-radius diagram together with various theoretical planetary models with different multilayer interior compositions \citep{2016ApJ...819..127Z}. The comparison of the loci of TOI-1470\,b and c with these models indicates that both planets are consistent with a volatile-rich composition and are mini-Neptunes rather than rocky super-Earths, consistent with their location on the high-radius regime of the planetary radius valley. \cite{2022Sci...377.1211L} reported that the bimodal distribution of the radii of small transiting exoplanets is also seen in the densities of the planets. They reported that only two compositions are predominant among small exoplanets: One planetary group (typically smaller radii and high density) is consistent with a purely rocky composition, and a second group (typically larger radii up to 2.5\,R$_{\oplus}$ and lower density) is likely composed of 50\,\%~rock and 50\,\%~water. Planets with even larger radii are expected to have H/He envelopes in addition to water-rich layers. TOI-1470\,b, the innermost planet, matches the second group (50\,\%~Earth-like rocky core and 50\% $\rm H_{2}O$ layer), while the location of TOI-1470\,c is better reproduced by the 100\%~water model with a thin or nonexistent H$_2$ envelope by mass.

We can use the rotation period to estimate the X-ray luminosity using the \citet{2011ApJ...743...48W} relation between X-ray emission and stellar rotation. Then we can use this number to estimate the extreme-UV (100-920~\AA) flux using the relations from \citet{sanz22}. For TOI-1470, we obtain $L_{\rm X}=9\times 10^{27}$\,erg\,s$^{-1}$ in the 5--100~\AA\ spectral range, and $L_{\rm EUV}=5\times 10^{28}$\,erg\,s$^{-1}$. The range 100--504~\AA\, that is relevant for He ionization has $L_{\rm EUV He}=3\times 10^{28}$\,erg\,s$^{-1}$. We estimate a mass-loss rate in the atmosphere of TOI-1470~b \citep[see][and references therein]{2011A&A...532A...6S} of $9\times 10^{10}$\,g\,s$^{-1}$, or 0.47\,M$_{\oplus}$\,Gyr$^{-1}$, and a much lower value of $8\times 10^{9}$\,g\,s$^{-1}$ in the case of TOI-1470~c.

Following the metrics proposed by \citep{2018PASP..130k4401K} to identify the \textit{TESS} transiting planets that are most amenable for atmospheric characterization via transmission spectroscopy with the James Webb Space Telescope (\textit{JWST}), we derived TSM = 55.5 and 37.6 for TOI-1470\,b and c, respectively, where TSM stands for the metric for transmission spectroscopy. It depends on the stellar brightness, the host star temperature, and the planetary properties, including the planet equilibrium and dayside temperatures. As a rule of thumb, for a given host star, a greater radius and higher temperature of the planet imply a higher TSM and a higher probability of detecting the planetary atmosphere. The TSM of both TOI-1470\,b and c is below the threshold metric values defined for their respective planetary categories (Earth-size and small sub-Neptune) by \cite{2018PASP..130k4401K}. This indicates that the two planets are poor targets for the study of their atmospheres, if they have any. An atmospheric characterization for TOI-1470\,b may not be very challenging for the JWST, however.

\section{Summary}
\label{sec:summary}

We presented the validation of TOI-1470\,b and the discovery of TOI-1470\,c, which orbit an M1.5V star. We used photometric \textit{TESS}, MuSCAT, MuSCAT2, and MuSCAT3 data and spectroscopic CARMENES observations. The inner planet was previously announced by \textit{TESS} as a transiting-planet candidate, but the outer planet has been discovered here in a spectroscopic and a photometric analysis. The intrinsic stellar variability was analyzed using the spectroscopic activity indicators provided by CARMENES and the photometric monitoring by ASAS-SN, TJO, OSN, and \textit{TESS}. Our findings suggest that TOI-1470 is a rather quiet star with a stellar rotation period of $29\pm 3$\,d.

The joint analysis of the \textit{TESS}, MuSCAT, MuSCAT2, MuSCAT3, and CARMENES data yields orbital periods of $2.527093^{+0.000004}_{-0.000003}$ and  $18.08816^{+0.00006}_{-0.00008}$\,d, masses of $7.32^{+1.21}_{-1.24}$ and $7.24^{+2.87}_{-2.77}$\,M$_{\oplus}$, and radii of $2.18^{+0.04}_{-0.04}$ and $2.47^{+0.02}_{-0.02}$\,R$_{\oplus}$ for the transiting planets TOI-1470\,b and TOI-1470\,c, respectively. The two planets are placed on the same side of the small-planet radius valley of M dwarfs, and the composition of the inner planet is compatible with a 50\% rocky world, whereas the outer planet is a water-rich world. Therefore, TOI-1470\,b and c are very interesting targets because the nature, formation, and evolution of sub-Neptune-sized planets are still open questions.


\begin{acknowledgements}

We thank the anonymous referee for helpful comments and suggestions, which helped to improve the manuscript. CARMENES is an instrument at the Centro Astron\'omico Hispano en Andaluc\'ia (CAHA) at Calar Alto (Almer\'{\i}a, Spain), operated jointly by the Junta de Andaluc\'ia and the Instituto de Astrof\'isica de Andaluc\'ia (CSIC). CARMENES was funded by the German Max-Planck- Gesellschaft (MPG), the Spanish Consejo Superior de Investigaciones Cient\'ificas (CSIC), the European Union through FEDER/ERF funds, and the members of the CARMENES Consortium (Max-Planck-Institut f\"ur Astronomie, Instituto de Astrof\'isica de Andaluc\'ia, Landessternwarte Ko\"onigstuhl, Institut de Ci\` encies de l'Espai, Insitut f\"ur Astrophysik G\"ottingen, Universidad, Complutense de Madrid, Th\"uringer Landessternwarte Tautenburg, Instituto de Astrof\'isica de Canarias, Hamburger Sternwarte, Centro de Astrobiolog\'ia and Centro Astron\'omico Hispano en Andaluc\'ia), with additional contributions by the Spanish Ministry of Economy, the state of Baden-W\"uttemberg, the Deutsche Forschungsge-meinschaft (DFG) through the Major Research Instrumentation Programme and Research Unit FOR2544 ``Blue Planets around Red Stars'', the Klaus Tschira Foundation, and by the Junta de Andaluc\'ia. This work was based on data from the CARMENES data archive at CAB (CSIC-INTA). 
This research has made use of the NASA Exoplanet Archive, which is operated by the California Institute of Technology, under contract with the National Aeronautics and Space Administration under the Exoplanet Exploration Program. Funding for the \textit{TESS} mission is provided by NASA’s Science Mission Directorate. This paper includes data collected by the TESS mission that are publicly available from the Mikulski Archive for Space Telescopes. We acknowledge the use of public TESS data from pipelines at the \textit{TESS} Science Office and at the \textit{TESS} Science Processing Operations Center. Resources supporting this work were provided by the NASA High-End Computing (HEC) Program through the NASA Advanced Supercomputing (NAS) Division at Ames Research Center for the production of the SPOC data products.  
This article is based on observations made with the MuSCAT2 instrument, developed by ABC, at Telescopio Carlos S\'anchez operated on the island of Tenerife by the IAC in the Spanish Observatorio del Teide. This paper is based on observations made with the MuSCAT3 instrument, developed by the Astrobiology Center and under financial supports by JSPS KAKENHI (JP18H05439) and JST PRESTO (JPMJPR1775), at Faulkes Telescope North on Maui, HI, operated by the Las Cumbres Observatory. Data were partly collected with the 1.50\,m telescope at the Observatorio de Sierra Nevada operated by the Instituto de Astrof\'\i fica de Andaluc\'\i a (IAA-CSIC). The Joan Or\'o Telescope (TJO) of the Montsec Observatory (OdM) is owned by the Generalitat de Catalunya and operated by the Institute for Space Studies of Catalonia (IEEC). We acknowledge the telescope operators from Observatori del Montsec, Observatorio de Sierra Nevada, and Centro Astron\'omico Hispano en Andaluc\'ia (CAHA) at Calar Alto. 
We acknowledge financial support from the Agencia Estatal de Investigaci\'on (AEI/10.13039/501100011033) of the Ministerio de Ciencia e Innovaci\'on and the ERDF ``A way of making Europe'' through projects 
PID2019-109522GB-C5[1:4],   
PID2019-107061GB-C64, and   
PID2019-110689RB-100,       
and the Centre of Excellence ``Severo Ochoa'' and ``Mar\'ia de Maeztu'' awards to the Instituto de Astrof\'isica de Canarias (CEX2019-000920-S), Instituto de Astrof\'isica de Andaluc\'ia (SEV-2017-0709), and Centro de Astrobiolog\'ia (MDM-2017-0737); 
the Generalitat de Catalunya/CERCA programme;
MEXT/JSPS KAKENHI through grants 15H02063, JP17H04574, JP18H05439, JP18H05442, JP20J21872, JP21K20376, and 22000005;
and JST CREST through grant JPMJCR1761.

\end{acknowledgements}

%
%

\bibliographystyle{aa} 
\bibliography{EGonzalez_toi1470.bib} 


\begin{appendix} 
\onecolumn

\section{Figures}

\begin{figure*}[!h]
        \centering
        \includegraphics[width=\textwidth]{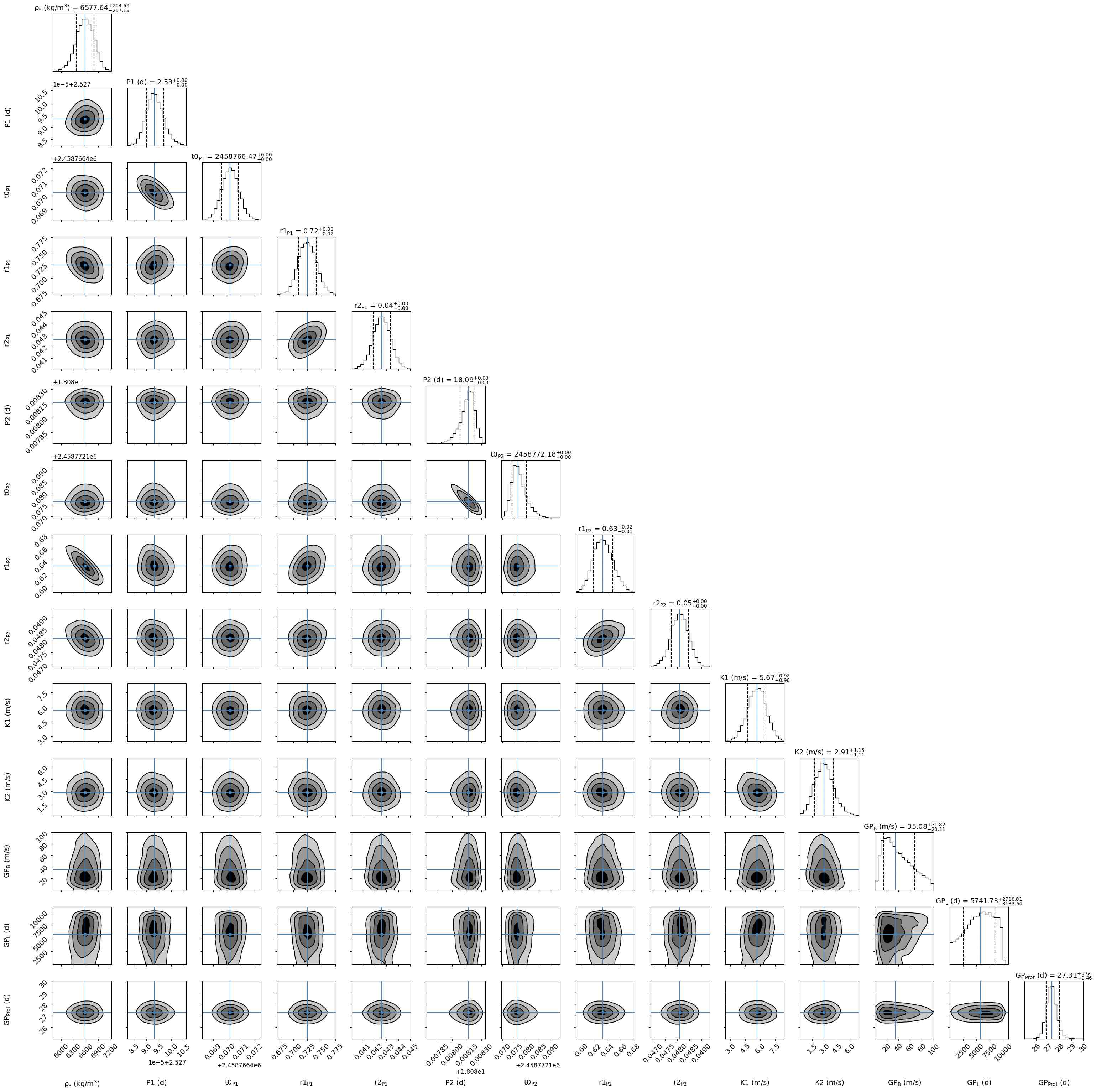}
        \caption{Posterior distributions of some of the planetary parameters of the TOI-1470 system we fit as obtained from the combined photometric and spectroscopic fit. The vertical dashed lines indicate the 16, 50, and 84\,\%~quantiles that were used to define the optimal values and their associated 1$\sigma$ uncertainty. The solid blue lines stand for the median values of each fit parameter. 
        }
        \label{fig:toi1470_cornerplot}
\end{figure*}

\clearpage

\newpage
\clearpage

\section{Tables}

\begin{table*}[!h]
\caption{\label{tab:toi1470_rv_act_data} TOI-1470 RV data from the CARMENES observations.}
\centering  
        
\begin{tabular}{l c c}
\hline
\hline
\noalign{\smallskip}
BJD & RV$_{\rm VIS}$  & $\delta$RV$_{\rm VIS}$ \\
(d)  & (m\,s$^{-1}$) & (m\,s$^{-1}$ )\\
\noalign{\smallskip}
\hline
\noalign{\smallskip}
2459003.6440 & $13.22$ & $ 5.37$ \\ 
2459010.6460 & $-4.81$ & $ 4.28$ \\ 
2459011.6420 & $-6.69$ & $ 4.63$ \\ 
2459013.5885 & $20.72$ & $ 6.36$ \\ 
2459015.6361 & $ 0.81$ & $ 2.88$ \\ 
2459018.6327 & $-0.78$ & $ 3.23$ \\ 
2459021.6275 & $ 6.04$ & $ 2.56$ \\ 
2459034.5910 & $-11.48$ & $ 2.42$ \\ 
2459040.6441 & $ 2.35$ & $ 2.51$ \\ 
2459041.5915 & $ 1.18$ & $ 2.80$ \\ 
2459049.6444 & $ 2.45$ & $ 2.35$ \\ 
2459050.5127 & $ 2.37$ & $ 2.70$ \\ 
2459054.6157 & $ 3.17$ & $ 2.27$ \\ 
2459055.6063 & $ 8.32$ & $ 2.26$ \\ 
2459056.6004 & $11.82$ & $ 2.79$ \\ 
2459058.6124 & $12.14$ & $ 2.96$ \\ 
2459059.6324 & $ 2.69$ & $ 2.69$ \\ 
2459061.6416 & $ 7.96$ & $ 2.39$ \\ 
2459063.6116 & $-1.41$ & $ 2.89$ \\ 
2459151.4357 & $-1.70$ & $ 2.54$ \\ 
2459152.4304 & $ 3.14$ & $ 2.34$ \\ 
2459156.4234 & $-12.06$ & $ 2.58$ \\ 
2459161.4262 & $-3.67$ & $ 3.50$ \\ 
2459162.4567 & $14.42$ & $10.34$ \\ 
2459163.4512 & $ 2.43$ & $ 4.46$ \\ 
2459171.3904 & $-7.29$ & $ 2.63$ \\ 
2459172.4156 & $-5.09$ & $ 2.56$ \\ 
2459172.4384 & $-2.09$ & $ 2.98$ \\ 
2459173.3859 & $-14.00$ & $ 3.55$ \\ 
2459174.3820 & $ 0.90$ & $ 4.72$ \\ 
2459175.3895 & $-0.04$ & $ 3.17$ \\ 
2459176.3916 & $-13.62$ & $ 2.69$ \\ 
2459177.3674 & $-5.57$ & $ 2.31$ \\ 
2459178.3640 & $-6.33$ & $ 3.65$ \\ 
2459186.4303 & $ 1.73$ & $ 3.78$ \\ 
2459187.3513 & $ 3.40$ & $ 4.09$ \\ 
2459193.3457 & $ 9.63$ & $ 5.36$ \\ 
2459196.3366 & $-1.78$ & $ 2.98$ \\ 
2459197.3874 & $-2.53$ & $ 4.36$ \\ 
2459215.3879 & $16.36$ & $ 4.84$ \\ 
2459218.3764 & $11.22$ & $ 4.87$ \\ 
2459219.3118 & $-4.64$ & $ 6.16$ \\ 
2459231.3085 & $-1.94$ & $ 2.80$ \\ 
2459232.3073 & $-11.54$ & $ 2.29$ \\ 
\\
\hline
\end{tabular}
\end{table*}

\begin{table*}[!ht]
        \centering
        \caption{Priors used for the joint LCs and RVs fit of TOI-1470.}
        \label{tab:1470_priors_details}
        \begin{tabular}{l c c r}

                \hline
                \hline
                \noalign{\smallskip}
                Parameter & Prior & Unit & Description \\
                \noalign{\smallskip}    
                \hline  
                \noalign{\smallskip}
                
                \multicolumn{4}{c}{\textit{Stellar parameter}} \\
                \noalign{\smallskip}
                
                $\rho_{\star}$                                          & $\mathcal{N}$(6.413, 0.279) & g\,cm$^{-3}$ & Stellar density\\
                
                \noalign{\smallskip}
                \multicolumn{4}{c}{\textit{Photometric parameters}} \\
                \noalign{\smallskip}
                
                $\mu_{\rm TESS_{S17,S18,S24,S58}}$                                              & $\mathcal{N}$(0, 0.1) & & The offset relative flux for \textit{TESS}\\
                $\mu_{\rm MuSCAT_{zs,r,i, g}}$                                                  & $\mathcal{N}$(0, 0.1) &   & The offset relative flux for \textit{TESS}\\

                $\sigma_{\rm TESS_{S17,S18,S24,S58}}$           & $\mathcal{L} \mathcal{U}$(10$^{-6}$, 0.04)  &   & A jitter added in quadrature to the error bars of instrument\\
                $\sigma_{\rm MuSCAT_{zs,r,i, g}}$               & $\mathcal{L} \mathcal{U}$(10$^{-6}$, 0.01) & & A jitter added in quadrature to the error bars of instrument\\                    
                
                $q1_{\rm TESS_{S17,S18,S24,S58}}$                                                       & $\mathcal{N}$(0.286, 0.009)     &       ...                     & Limb-darkening for photometric instrument\\
                $q2_{\rm TESS_{S17,S18,S24,S58}}$                                                       & $\mathcal{N}$(0.270, 0.008) &   ...             & Limb-darkening for photometric instrument\\
                
                $q1_{\rm MuSCA\, r}$                                                    & $\mathcal{N}$(0.68, 0.02)       &       ...                     & Limb-darkening for photometric instrument\\
                $q2_{\rm MuSCAT\, r}$                                                   & $\mathcal{N}$(0.316, 0.09) &    ...             & Limb-darkening for photometric instrument\\

                $q1_{\rm MuSCAT\, i}$                                                   & $\mathcal{N}$(0.324, 0.009)     &       ...                     & Limb-darkening for photometric instrument\\
                $q2_{\rm MuSCAT\, i}$                                                   & $\mathcal{N}$(0.283, 0.009) &   ...             & Limb-darkening for photometric instrument\\            
                
                $q1_{\rm MuSCAT\, zs}$                                                  & $\mathcal{N}$(0.209, 0.007)     &       ...                     & Limb-darkening for photometric instrument\\
                $q2_{\rm MuSCAT\, zs}$                                                  & $\mathcal{N}$(0.260, 0.009) &   ...             & Limb-darkening for photometric instrument\\    
                
                $q1_{\rm MuSCAT\, g}$                                                   & $\mathcal{N}$(0.79, 0.02)       &       ...                     & Limb-darkening for photometric instrument\\
                $q2_{\rm MuSCAT\, g}$                                                   & $\mathcal{N}$(0.321, 0.008) &   ...             & Limb-darkening for photometric instrument\\    
                
                $D_{\rm TESS_{S17,S18,S24}}$                                                            & 1 (fixed)                                       &       ...             & The dilution factor for the photometric instrument\\
                 $D_{\rm MuSCAT_{zs,r,i, g}}$                                                           & 1 (fixed)                                       &       ...             & The dilution factor for the photometric instrument\\            
                 
                \noalign{\smallskip}
                \multicolumn{4}{c}{\textit{RV parameters}} \\
                \noalign{\smallskip}
                
                $\gamma$                                                                &  $\mathcal{U}$(-10, 10) & m\,$\rm s^{-1}$ & RV zero point for CARMENES\\
                $\sigma$                                                                        & $\mathcal{L} \mathcal{U}$(0.001, 5) & m\,$\rm s^{-1}$ & A jitter added in quadrature \\
                
                \noalign{\smallskip}                                                        
                \multicolumn{4}{c}{\textit{GP parameters}} \\
                \noalign{\smallskip}            
                
                $B_{\rm GP,RV}$                             & $\mathcal{U}$(0.01, 100)   & m\,$\rm s^{-1}$ & Related to the amplitude of the GP for the RVs\\
                $L_{\rm GP,RV}$                                 & $\mathcal{U}$(10, 10$^{4}$) & d & Related with the timescale for the amplitude-modulation of the GP\\
                $P_{\rm rot, GP,RV}$                                            & $\mathcal{U}$(20, 50) & d & Period of the quasi-periodic kernel\\
                
                \noalign{\smallskip}                                                        
                \multicolumn{4}{c}{\textit{Planet $b$ parameters} }\\   
                \noalign{\smallskip}
                
                $P$                                                                             & $\mathcal{N}$ (2.527, 0.01)     & d             & Period \\
                $t_0$ (BJD\text{--}2,457,000)           & $\mathcal{N}$ (1766.47, 0.01)           & d             &  Time of periastron passage\\
                $e$                                                                             & 0 (fixed)                                                                                       & ...     & Orbital eccentricity \\
                $\omega$                                                                &       90 (fixed)                                                                         & deg     & Periastron angle \\
                $K$                                                                                     &  $\mathcal{U}$ (0, 10)                                  &        m\,$\rm s^{-1}$          & RV semi-amplitude  \\   
                $r_1$                                                                           &  $\mathcal{U}$ (0, 1)                                   & ...        & Parameterization for $p$ and $b$  \\   
                $r_2$                                                                           &  $\mathcal{U}$ (0, 1)                                   & ...        & Parameterization for $p$ and $b$    \\   
                
                \noalign{\smallskip}                                                        
                \multicolumn{4}{c}{\textit{Planet $c$ parameters} }\\   
                \noalign{\smallskip}                                                        
                
                $P$                                                                             & $\mathcal{N}$ (18.08, 0.01)             & d             & Period \\
                $t_0$ (BJD\text{--}2,457,000)   & $\mathcal{N}$ (1772.17, 0.1)            & d             &  Time of periastron passage\\
                $e$                                                                             & 0 (fixed)                                                                                               & ... & Orbital eccentricity\\
                $\omega$                                                                &       90 (fixed)                                                                                 & deg     & Periastron angle\\
                $K$                                                                                     &  $\mathcal{U}$ (0, 10)          &         m\,$\rm s^{-1}$        & RV semi-amplitude\\   
                $r_1$                                                                           &  $\mathcal{U}$ (0, 1)   &  ...       & Parameterization for $p$ and $b$  \\   
                $r_2$                                                                           &  $\mathcal{U}$ (0, 1)   &  ...       & Parameterization for $p$ and $b$    \\

                \noalign{\smallskip}    
                \hline  
                \noalign{\smallskip}
        \end{tabular}
        \tablefoot{The prior labels of $\mathcal{N}$, $\mathcal{U}$, and $\mathcal{L}$ $\mathcal{U}$ represent the normal, uniform, and log-uniform distribution, respectively. The error on the density of the star comes from the stellar mass and radius errors. The upper limit on the photometric jitter term corresponds to ten times the error bars of the photometric data.}
\end{table*}

\begin{table}[]
        \renewcommand{\arraystretch}{1.3}
                \centering
                \caption{Final adopted parameters for the TOI-1470 system.}
                \label{tab:toi1470bc_params_from_joit-fit_rest}
                \begin{tabular}{l c c }
                        
                        \hline
                        \hline
                        \noalign{\smallskip}
                        
                        Parameter & Value                &   \\
                        \hline
                        \noalign{\smallskip}

                        \noalign{\smallskip}
                        \multicolumn{3}{c}{\textit{Photometric parameters} }\\
                        \noalign{\smallskip}
                        $\mu_{\rm TESS, S17}$ (ppm)                     &    \multicolumn{2}{c}{$-3.2^{+32.9}_{-32.6}$} \\
                        $\mu_{\rm TESS, S18}$ (ppm)                     &    \multicolumn{2}{c}{$-8.6^{+33.8}_{-31.9}$} \\
                        $\mu_{\rm TESS, S24}$ (ppm)                     &    \multicolumn{2}{c}{$2.7^{+32.2}_{-32.2}$} \\
            $\mu_{\rm TESS, S58}$ (ppm)                         &    \multicolumn{2}{c}{$-6.3^{+17.5}_{-16.6}$} \\
                        
                        $\mu_{\rm MuSCAT zs}$ (ppm)                     &    \multicolumn{2}{c}{$4.6^{+45.9}_{-46.5}$} \\
                        $\mu_{\rm MuSCAT r}$ (ppm)                      &    \multicolumn{2}{c}{$44.7^{+50.9}_{-52.1}$} \\
                        $\mu_{\rm MuSCAT2 zs}$ (ppm)                    &    \multicolumn{2}{c}{$-14.8^{+51.5}_{-53.7}$} \\
                        $\mu_{\rm MuSCAT2 i}$ (ppm)                     &    \multicolumn{2}{c}{$501.2^{+61.9}_{-61.3}$} \\
                        $\mu_{\rm MuSCAT3 zs}$ (ppm)                    &    \multicolumn{2}{c}{$252.8^{+38.8}_{-37.7}$} \\
                        $\mu_{\rm MuSCAT3 i}$ (ppm)                     &    \multicolumn{2}{c}{$-20.6^{+39.1}_{-37.7}$} \\
                        $\mu_{\rm MuSCAT3 g}$ (ppm)                     &    \multicolumn{2}{c}{$-14.0^{+58.1}_{-57.4}$} \\
                        
                        $\sigma_{\rm TESS, S17}$                                                                                &    \multicolumn{2}{c}{$0.0001^{+0.0041}_{-0.0001}$} \\
                        $\sigma_{\rm TESS, S18}$                                                                                &    \multicolumn{2}{c}{$0.0004^{+0.0076}_{-0.0004}$} \\
                        $\sigma_{\rm TESS, S24}$                                                                                &    \multicolumn{2}{c}{$0.0003^{+0.0059}_{-0.0003}$} \\
            $\sigma_{\rm TESS, S58}$                                                                            &    \multicolumn{2}{c}{$0.0002^{+0.0031}_{-0.0002}$} \\

                        $\sigma_{\rm MuSCAT zs}$                                                        &    \multicolumn{2}{c}{$0.0001^{+0.0012}_{-0.0001}$} \\
                        $\sigma_{\rm  MuSCAT r}$                                                                &    \multicolumn{2}{c}{$0.0001^{+0.0027}_{-0.0001}$} \\
                        $\sigma_{\rm  MuSCAT2 zs}$                                                      &    \multicolumn{2}{c}{$0.0002^{+0.0039}_{-0.0003}$} \\  
                        $\sigma_{\rm  MuSCAT2 i}$                                                       &    \multicolumn{2}{c}{$0.0001^{+0.0027}_{-0.0001}$} \\                  
                        $\sigma_{\rm  MuSCAT3 zs}$                                                      &    \multicolumn{2}{c}{$0.0001^{+0.0015}_{-0.0001}$} \\          
                        $\sigma_{\rm  MuSCAT3 i}$                                                       &    \multicolumn{2}{c}{$0.0001^{+0.0011}_{-0.0001}$} \\  
                        $\sigma_{\rm  MuSCAT3 g}$                                                       &    \multicolumn{2}{c}{$0.0001^{+0.0015}_{-0.0001}$} \\

                        $q1_{\rm TESS}$                                                                         &    \multicolumn{2}{c}{$0.287^{+0.008}_{-0.008}$} \\
                        $q2_{\rm TESS}$                                                                         &    \multicolumn{2}{c}{$0.269^{+0.007}_{-0.007}$} \\
                        
                        $q1_{\rm MuSCAT z}$                                                                     &    \multicolumn{2}{c}{$0.208^{+0.006}_{-0.006}$} \\
                        $q2_{\rm MuSCAT z}$                                                                     &    \multicolumn{2}{c}{$0.259^{+0.008}_{-0.007}$} \\             
                        
                        $q1_{\rm MuSCAT r}$                                                                     &    \multicolumn{2}{c}{$0.68^{+0.02}_{-0.02}$} \\
                        $q2_{\rm MuSCAT r}$                                                                     &    \multicolumn{2}{c}{$0.314^{+0.008}_{-0.007}$} \\             
                        
                        $q1_{\rm MuSCAT i}$                                                                     &    \multicolumn{2}{c}{$0.328^{+0.008}_{-0.007}$} \\
                        $q2_{\rm MuSCAT i}$                                                                     &    \multicolumn{2}{c}{$0.284^{+0.007}_{-0.008}$} \\     
                        
                        $q1_{\rm MuSCAT g}$                                                                     &    \multicolumn{2}{c}{$0.79^{+0.02}_{-0.02}$} \\
                        $q2_{\rm MuSCAT g}$                                                                     &    \multicolumn{2}{c}{$0.320^{+0.008}_{-0.008}$} \\                             
                        
                        \noalign{\smallskip}
                        \multicolumn{3}{c}{\textit{RV parameters}} \\
                        \noalign{\smallskip}
                        $\gamma$ (m\,s$^{-1}$)                  &    \multicolumn{2}{c}{$-0.36^{+3.41}_{-3.65}$} \\
                        $\sigma$ (m\,s$^{-1}$)                          &    \multicolumn{2}{c}{$2.8^{+0.8}_{-0.9}$} \\
                        
                        \noalign{\smallskip}
                        \multicolumn{3}{c}{\textit{GP hyperparameters}} \\
                        \noalign{\smallskip}
                        $B_{\rm GP,RV}$ (m\,s$^{-1}$)                                                     &  \multicolumn{2}{c}{$35.08^{+31.8}_{-20.1}$} \\
                        $L_{\rm GP,RV}$ (d)                                                                     &   \multicolumn{2}{c}{$5741^{+2719}_{-3184}$} \\
                        $P_{\rm rot, GP,RV}$ (d)                                                                &   \multicolumn{2}{c}{$27.3^{+0.6}_{-0.5}$} \\
                        
                        \noalign{\smallskip}    
                        \noalign{\smallskip}
                        \multicolumn{3}{c}{\textit{Stellar parameters}}\\
                        
                        $\rho_{\rm \star}$ (g $\rm cm^{-3}$)            &                                          \multicolumn{2}{c}{$6.57^{+0.21}_{-0.22}$} \\
                        
                        \noalign{\smallskip}
                        \hline
                        \noalign{\smallskip}

                \end{tabular}
\end{table}

\longtab[0]{
        \begin{landscape}
                \begin{scriptsize}
                        \begin{longtable}{lcc}
                                \caption{\label{tab:toi1470_rv_act_data} TOI-1470 RV data from the CARMENES observations.} \\
                                \hline\hline
                                \noalign{\smallskip}
                                BJD & RV$_{\rm VIS}$  & eRV$_{\rm VIS}$ \\
                                (d)  & (m\,s$^{-1}$) & (m\,s$^{-1}$ )\\
                                \noalign{\smallskip}
                                \hline
                                \noalign{\smallskip}
                                \endfirsthead
                                
                                \caption{continued.}  \\
                                
                                \hline\hline
                                \noalign{\smallskip}
                                BJD & RV$_{\rm VIS}$  & eRV$_{\rm VIS}$\\
                                (d)  & (m\,s$^{-1}$) & (m\,s$^{-1}$)\\
                                \noalign{\smallskip}
                                \hline
                                \endhead
                                
                                \hline
                                \endfoot
                                
                        \end{longtable}
                \end{scriptsize}
        \end{landscape}
        
}

\end{appendix}

\end{document}